\overfullrule=0pt
\input harvmac
\input tikz.tex
\input mathdots 

\def\a{{\alpha}}

\def\l{{\lambda}}

\def\b{{\beta}}

\def\g{{\gamma}}

\def\d{{\delta}}
\def\e{{\epsilon}}
\def\s{{\sigma}}

\def\half{{1\over 2}}
\def\p{{\partial}}

\def\t{{\theta}}

\def\bar{\overline}
\def\({\left(}
\def\){\right)}

\def\frac#1#2{{#1 \over #2}}
\def\ap{\alpha'}
\def\br{\hfill\break}

\newread\instream \openin\instream= labeldefs.tmp
\ifeof\instream \message{No labels in advance yet. Wait till next pass.}
\else \closein\instream \input labeldefs.tmp
\fi
\writedefs

\font\smallrm=cmr8
\input epsf
\def\figin{\epsfcheck\figin}\def\figins{\epsfcheck\figins}
\def\epsfcheck{\ifx\epsfbox\UnDeFiNeD
\message{(NO epsf.tex, FIGURES WILL BE IGNORED)}
\gdef\figin##1{\vskip2in}\gdef\figins##1{\hskip.5in}
\else\message{(FIGURES WILL BE INCLUDED)}%
\gdef\figin##1{##1}\gdef\figins##1{##1}\fi}
\def\DefWarn#1{}
\def\figinsert{\goodbreak\midinsert}
\def\ifig#1#2#3{\DefWarn#1\xdef#1{Fig.~\the\figno}
\writedef{#1\leftbracket fig.\noexpand~\the\figno}%
\figinsert\figin{\centerline{#3}}\medskip\centerline{\vbox{\baselineskip12pt
\advance\hsize by -1truein\noindent\footnotefont\centerline{{\bf
Fig.~\the\figno}\ #2}}}
\smallskip\endinsert\noindent\global\advance\figno by1}

\def\DefWarn#1{}
\def\tikzcaption#1#2{\DefWarn#1\xdef#1{Fig.~\the\figno}{
\bigskip
\leftskip=40pt\rightskip=40pt\noindent\font\smallrm=cmr6
{{\bf Fig.~\the\figno}\ #2}
\smallskip\bigskip
\global\advance\figno by1 \par}}


\Title{\vbox{\rightline{AEI--2011--024}
\rightline{MPP--2011--46}
\rightline{NSF--ITP--11--058}\vskip-1.75cm}}
{\vbox{\centerline{Explicit BCJ Numerators from Pure Spinors}}}
\centerline{Carlos R. Mafra$^{a,b}$,
Oliver Schlotterer$^{b,c}$, and
Stephan Stieberger$^{b,c}$}
\bigskip\medskip
\centerline{\it $^a$ Max--Planck--Institut f\"ur Gravitationsphysik, Albert--Einstein--Institut,
14476 Potsdam, Germany}
\vskip4pt
\centerline{\it $^b$ Kavli Institute for Theoretical Physics, University of California, 
Santa Barbara, CA 93106, USA}
\vskip4pt
\centerline{\it $^c$ Max--Planck--Institut f\"ur Physik, Werner--Heisenberg--Institut, 
80805 M\"unchen, Germany}
\medskip
\centerline{\tt E-mails: crmafra@aei.mpg.de, olivers@mppmu.mpg.de,}
\centerline{\tt stephan.stieberger@mpp.mpg.de}

\bigskip\medskip
\centerline{\bf Abstract}
\vskip1pt
\noindent

We derive local kinematic numerators for gauge theory tree amplitudes which manifestly satisfy
Jacobi identities analogous to color factors. They naturally emerge from the low energy limit of superstring amplitudes
computed with the pure spinor formalism.
The manifestation of the color--kinematics duality is a consequence of the superstring computation involving no more 
than $(n-2)!$ kinematic factors for the full color dressed $n$ point amplitude. 
The bosonic part of these results describe gluon scattering independent on the number of supersymmetries and captures 
any N$^k$MHV helicity configuration after dimensional reduction to $D=4$ dimensions.

\Date{}
\noindent

\goodbreak


\lref\BernUE{
  Z.~Bern, J.J.M.~Carrasco, H.~Johansson,
  ``Perturbative Quantum Gravity as a Double Copy of Gauge Theory,''
Phys.\ Rev.\ Lett.\  {\bf 105}, 061602 (2010).
[arXiv:1004.0476 [hep-th]].
}

\lref\StieOpr{
  D.~Oprisa and S.~Stieberger,
  ``Six gluon open superstring disk amplitude, multiple hypergeometric  series
  and Euler-Zagier sums,''
  arXiv:hep-th/0509042.
}
\lref\tye{
  S.H.~Henry Tye and Y.~Zhang,
``Dual Identities inside the Gluon and the Graviton Scattering Amplitudes,''
  JHEP {\bf 1006}, 071 (2010)
  [arXiv:1003.1732 [hep-th]].
}
\lref\Medinas{
  R.~Medina, F.T.~Brandt and F.R.~Machado,
  ``The open superstring 5-point amplitude revisited,''
  JHEP {\bf 0207}, 071 (2002)
  [arXiv:hep-th/0208121]
\semi
  L.A.~Barreiro and R.~Medina,
  ``5-field terms in the open superstring effective action,''
  JHEP {\bf 0503}, 055 (2005)
  [arXiv:hep-th/0503182].
}
\lref\StieSusy{
  S.~Stieberger, T.R.~Taylor,
  ``Amplitude for N-Gluon Superstring Scattering,''
Phys.\ Rev.\ Lett.\  {\bf 97}, 211601 (2006).
[hep-th/0607184];
``Multi-Gluon Scattering in Open Superstring Theory,''
Phys.\ Rev.\  {\bf D74}, 126007 (2006).
[hep-th/0609175];
  ``Supersymmetry Relations and MHV Amplitudes in Superstring Theory,''
Nucl.\ Phys.\  {\bf B793}, 83-113 (2008).
[arXiv:0708.0574 [hep-th]];
 ``Complete Six-Gluon Disk Amplitude in Superstring Theory,''
Nucl.\ Phys.\  {\bf B801}, 128-152 (2008).
[arXiv:0711.4354 [hep-th]].
}

\lref\ParkeTaylor{
  S.J.~Parke, T.R.~Taylor,
  ``An Amplitude for $n$ Gluon Scattering,''
Phys.\ Rev.\ Lett.\  {\bf 56}, 2459 (1986).
}
\lref\FTAmps{
  C.R.~Mafra,
  ``Towards Field Theory Amplitudes From the Cohomology of Pure Spinor
  Superspace,''
  JHEP {\bf 1011}, 096 (2010)
  [arXiv:1007.3639 [hep-th]].
}
\lref\wittentwistor{
  E.~Witten,
  ``Twistor - Like Transform In Ten-Dimensions,''
  Nucl.\ Phys.\  B {\bf 266}, 245 (1986).
}
\lref\psf{
  N.~Berkovits,
  ``Super-Poincare covariant quantization of the superstring,''
  JHEP {\bf 0004}, 018 (2000)
  [arXiv:hep-th/0001035].
}
\lref\thetaSYM{
  	J.P.~Harnad and S.~Shnider,
	``Constraints And Field Equations For Ten-Dimensional Superyang-Mills
  	Theory,''
  	Commun.\ Math.\ Phys.\  {\bf 106}, 183 (1986)
\semi
	P.A.~Grassi and L.~Tamassia,
        ``Vertex operators for closed superstrings,''
        JHEP {\bf 0407}, 071 (2004)
        [arXiv:hep-th/0405072].
}
\lref\tsimpis{
  G.~Policastro and D.~Tsimpis,
``$R^4$, purified,''
  Class.\ Quant.\ Grav.\  {\bf 23}, 4753 (2006)
  [arXiv:hep-th/0603165].
}
\lref\FivePt{
  C.R.~Mafra,
``Simplifying the Tree-level Superstring Massless Five-point Amplitude,''
  JHEP {\bf 1001}, 007 (2010)
  [arXiv:0909.5206 [hep-th]].
}
\lref\mafraids{
  C.R.~Mafra,
  ``Pure Spinor Superspace Identities for Massless Four-point Kinematic Factors,''
JHEP {\bf 0804}, 093 (2008).
[arXiv:0801.0580 [hep-th]].
}

\lref\MSSTFT{
  C.R.~Mafra, O.~Schlotterer, S.~Stieberger and D.~Tsimpis,
  ``A recursive formula for N-point SYM tree amplitudes,''
  arXiv:1012.3981 [hep-th].
}
\lref\MSST{
  C.R.~Mafra, O.~Schlotterer, S.~Stieberger and D.~Tsimpis,
  ``Six Open String Disk Amplitude in Pure Spinor Superspace,''
  Nucl.\ Phys.\  B {\bf 846}, 359 (2011)
  [arXiv:1011.0994 [hep-th]].
}
\lref\siegel{
  W.~Siegel,
  ``Classical Superstring Mechanics,''
Nucl.\ Phys.\  {\bf B263}, 93 (1986).
}
\lref\PSS{
  C.R.~Mafra,
  ``PSS: A FORM Program to Evaluate Pure Spinor Superspace Expressions,''
[arXiv:1007.4999 [hep-th]].
}
\lref\BCJ{
  Z.~Bern, J.J.M.~Carrasco, H.~Johansson,
  ``New Relations for Gauge-Theory Amplitudes,''
Phys.\ Rev.\  {\bf D78}, 085011 (2008).
[arXiv:0805.3993 [hep-ph]].
}
\lref\anomaly{
  N.~Berkovits and C.R.~Mafra,
  ``Some superstring amplitude computations with the non-minimal pure spinor
  formalism,''
  JHEP {\bf 0611}, 079 (2006)
  [arXiv:hep-th/0607187].
}
\lref\BG{
  F.A.~Berends, W.T.~Giele,
  ``Recursive Calculations for Processes with n Gluons,''
Nucl.\ Phys.\  {\bf B306}, 759 (1988).
}
\lref\FORM{
  J.A.M.~Vermaseren,
  ``New features of FORM,''
  arXiv:math-ph/0010025.
\semi
  M.~Tentyukov and J.A.M.~Vermaseren,
  ``The multithreaded version of FORM,''
  arXiv:hep-ph/0702279.
}
\lref\KK{
  R.~Kleiss, H.~Kuijf,
  ``Multi - Gluon Cross-sections And Five Jet Production At Hadron Colliders,''
Nucl.\ Phys.\  {\bf B312}, 616 (1989).
}
\lref\monodVanhove{
  N.E.J.~Bjerrum-Bohr, P.H.~Damgaard, P.~Vanhove,
  ``Minimal Basis for Gauge Theory Amplitudes,''
Phys.\ Rev.\ Lett.\  {\bf 103}, 161602 (2009).
[arXiv:0907.1425 [hep-th]].
}
\lref\monodStie{
  S.~Stieberger,
  ``Open \& Closed vs. Pure Open String Disk Amplitudes,''
[arXiv:0907.2211 [hep-th]].
}
\lref\ExtendedBCJs{
  N.E.J.~Bjerrum-Bohr, P.H.~Damgaard, T.~Sondergaard, P.~Vanhove,
  ``Monodromy and Jacobi-like Relations for Color-Ordered Amplitudes,''
JHEP {\bf 1006}, 003 (2010).
[arXiv:1003.2403 [hep-th]].
}
\lref\WIP{
C.R.~Mafra, O.~Schlotterer and S.~Stieberger, to appear
}
\lref\VamanP{
  D.~Vaman, Y.-P.~Yao,
  ``Constraints and Generalized Gauge Transformations on Tree-Level Gluon and Graviton Amplitudes,''
JHEP {\bf 1011}, 028 (2010).
[arXiv:1007.3475 [hep-th]].
}
\lref\ICTP{
  N.~Berkovits,
  ``ICTP lectures on covariant quantization of the superstring,''
[hep-th/0209059].
}
\lref\GGI{
  O.A.~Bedoya, N.~Berkovits,
  ``GGI Lectures on the Pure Spinor Formalism of the Superstring,''
[arXiv:0910.2254 [hep-th]].
}

\lref\Kiermaier{
M.~Kiermaier, Amplitudes 2010, Queen Mary, University of London, \br
 {\tt http://www.strings.ph.qmul.ac.uk/$\sim$theory/Amplitudes2010/Talks/MK2010.pdf}}
 

\lref\BBDSV{
 N.E.J.~Bjerrum-Bohr, P.H.~Damgaard, T.~Sondergaard and P.~Vanhove,
 ``The Momentum Kernel of Gauge and Gravity Theories,''
 JHEP {\bf 1101}, 001 (2011)
 [arXiv:1010.3933 [hep-th]].
}
\lref\HoweMF{
  P.S.~Howe,
  ``Pure spinors lines in superspace and ten-dimensional supersymmetric theories,''
Phys.\ Lett.\  {\bf B258}, 141-144 (1991).
}
\lref\NimaCatalan{
  N.~Arkani-Hamed, F.~Cachazo, C.~Cheung, J.~Kaplan,
  ``The S-Matrix in Twistor Space,''
JHEP {\bf 1003}, 110 (2010).
[arXiv:0903.2110 [hep-th]].
}

\lref\BernQN{
  Z.~Bern, J.~J.~Carrasco, L.~J.~Dixon, H.~Johansson, R.~Roiban,
  ``Amplitudes and Ultraviolet Behavior of N=8 Supergravity,''
[arXiv:1103.1848 [hep-th]].
}

\newsec{Introduction}

Bern, Carrasco and Johansson (BCJ) have introduced a parametrization of gauge theory scattering amplitudes 
such that all the kinematic factors obey an equivalent of the Jacobi identity for color factors \BCJ. This 
duality between color- and kinematic degrees of freedom is an excellent example for hidden simplicity and non-obvious 
harmony in scattering amplitudes. It plays a crucial role for taming the non-planar sector of SYM and for understanding gravity 
as the double copy of gauge theories \BernUE.

The BCJ organization scheme represents gauge theory amplitudes in terms of diagrams with cubic vertices only 
(in short: cubic diagrams). This amounts to writing the color-dressed $n$ point amplitude as follows:
\eqn \org{
{\cal A}_n = \sum_i {c_i  n_i \over \prod_{\alpha_i} s_{\alpha_i}} 
}
The $c_i$ denote color factors made of $n-2$ structure constants $f^{abc}$ of the gauge group, and their dual 
numerators $n_i$ are constructed in this work. Each $(n_i,c_i)$ pair multiplies $n-3$ propagators $s_{\alpha_i}^{-1}$ 
of a cubic $n$ point tree diagram.

The contribution of four point vertices in SYM fields to \org\ certainly contains less propagators and must be absorbed 
into the $n_i$ by multiplying with some ${ s_{\alpha_i} \over  s_{\alpha_i}}$ for compatibility with the pole structure. 
These contact terms introduce ambiguities in the parametrization above. At $n\geq 5$ points, a generic choice of assignment 
spoils the dual Jacobi identities for the $n_i$. Hence, contact terms have always been an obstacle in constructing color-dual 
BCJ numerators directly from the gauge theory. There exist Kawai--Lewellen--Tye (KLT) inspired expressions for $n_i$ in terms of color ordered gauge 
theory amplitudes \refs{\Kiermaier,\BBDSV} which do not exhibit manifest locality.
 
The approach used in this paper bypasses the contact term ambiguity because the new BRST cohomology organization of the string
amplitude discussed in \WIP\ naturally absorbs these contact terms. In \refs{\FivePt,\ExtendedBCJs} the contact term ambiguity 
was shown to arise from the double pole in the OPE of two integrated vertices in the field-theory limit of the string amplitude.
However, terms of this form are uniquely packaged inside BRST-covariant building blocks 
when the result of the tree-level string amplitude computed with the pure spinor formalism is recast into a form which
manifests its BRST properties \WIP.

The $n_i$ constructed this way can therefore be recycled from planar ${\cal N}=4$ SYM to the non-planar sector and used for ${\cal N}=8$ supergravity 
by means of the double copy construction, cf. \BernQN\ and references therein. Ultimately, they are helpful for studying the ultraviolet properties of gravity 
theories at higher loops.

Due to the pure spinor methods, the BCJ numerators are written in 
terms of ${\cal N}=1$ SYM superfields in $D=10$ dimensions. 
It is straightforward to dimensionally reduce the superfield components 
to $D=4$, and the bosonic parts describe gluon scattering independently on the existence of supersymmetries. The computation does not single out 
any particular helicity configuration and treats all N$^k$MHV amplitudes on the same footing.

\newsec{Basic facts of the Pure Spinor Formalism}

The explicit construction of BCJ numerators will be carried out in pure spinor superspace and
we show the fine structure of the $(n-2)!$ basic kinematics provided by the stringy computation of $n$-point 
amplitudes. For this purpose, this section briefly presents the tree-level framework of the pure spinor 
formalism \psf \foot{For
reviews of the pure spinor formalism, see \refs{\ICTP,\GGI}.}. In particular, we will make use of the pure spinor BRST cohomology
building blocks which are more carefully explained in \WIP.

The pure spinor formalism is a manifestly super-Poincare covariant approach to the superstring. Massless states of 
the open string sector are described by the ten-dimensional ${\cal N}=1$ super Yang-Mills superfields 
$[A_\a,A_m,W^\a,{\cal F}_{mn}]$ of \wittentwistor\ which encompass the gluon- polarization vector $e^m$ and the 
gluino wave function $\chi^{\alpha}$ and can be expanded in the Grassmann odd superspace coordinates $\theta^{\alpha}$.

They satisfy the following linearized equations of motion,
$$
D_\a A_\b + D_\b A_\a = \g^m_{\a\b} A_m,\quad
D_\a A_m = (\g_m W)_\a + k_m A_\a,
$$
\eqn\SYM{
D_\a{\cal F}_{mn} = 2k_{[m} (\g_{n]} W)_\a, \quad
D_\a W^{\b} = {1\over 4}(\g^{mn})^{\phantom{m}\b}_\a{\cal F}_{mn}
}
which imply the on-shell constraints $k_m e^m = k_m \gamma^m_{\alpha \beta} \chi^{\beta} = 0$ in components.

The scattering amplitudes of these massless states are obtained by computing the correlation function
\eqn\treepresc{
{\cal A}_n = \langle V^1(0)V^{(n-1)}(1)V^n(\infty) \int dz_2 U^2(z_2) 
{\ldots} \int dz_{(n-2)}U^{(n-2)}(z_{(n-2)})\rangle,
}
where $V^i$ and $U^i$ are vertex operators writen in terms of the SYM superfields
\eqn\vertices{
V^i = \l^\a A^i_\a(x,\t), \quad 
U^i = \p\t^\a A^i_\a + \Pi^m A^i_m + d_\a W^\a_i + \half {\cal F}_{mn}^i N^{mn}.
}
The bosonic ghost field $\l^\a(z)$ is a pure spinor satisfying $\l^\a \g^m_{\a\b} \l^\b = 0$ and
\eqn\susymom{
\Pi^m(z) = \p X^m + \half (\t\g^m \p\t), \quad d_\a(z) = \p_\a -\half(\g^m\t)_{\a}\p X_m 
-{1\over 8}(\g^m\t)_{\a}(\t\g_m\p\t)
}
are, respectively, the supersymmetric momentum and Green-Schwarz constraint. The Lorentz
currents for the pure spinor field are $N^{mn}(z) = \half (\l\g^{mn}w)$,
where $w_\a(z)$ is the conjugate momentum of $\l^\a(z)$ \psf. Their OPE's are easily
computed \refs{\ICTP,\siegel}.

The pure spinor BRST charge is defined by \psf
\eqn\BRSTc{
Q = \oint \l^\a(z)d_\a(z)
}
and satisfies $Q^2=0$. It can be shown that imposing $QV=0$ puts all superfields
on-shell, which also implies that the BRST variation of the integrated vertex $U(z)$ can be written
as $QU=\p V$ \ICTP\ (see also \HoweMF).

After using the OPE's to integrate out the conformal-weight one variables from the correlator
\treepresc, one is left with an expression containing only zero-modes of the form
\eqn\endres{
{\cal A}_n = \langle \l^\a\l^\b\l^\g f_{\a\b\g}^{i_1{\ldots} i_n}(\t,\a') \rangle.
}
In \endres, $f^{i_1{\ldots} i_n}_{\a\b\g}(\t,\a')$ is both a composite superfield in the labels $[i_1,{\ldots} i_n]$ of the external states 
and a function of the string scale $\a'$ satisfying $\l^\a\l^\b\l^\g\l^\d D_\d f_{\a\b\g}^{i_1{\ldots} i_n}(\t,\a')=0 $ due t BRST invariance. 
Its specific form in terms of the super Yang-Mills superfields $[A^i_\a,A^i_m,W^\a_i,{\cal F}_{mn}^i]$
follows from OPE contractions while its functional dependence on $\a'$
is determined by the momentum expansion of $n$-point Gaussian hypergeometric  functions \Medinas\ and multiple Gaussian hypergeometric functions \refs{\StieOpr,\StieSusy}.
The zero-mode integration is denoted by the pure spinor bracket $\langle {\ldots} \rangle$ \psf. It 
selects from the $\t-$expansion \refs{\thetaSYM,\tsimpis} 
of the enclosed superfields the unique
element in the cohomology of the pure spinor BRST operator at ghost-number three, and its tree-level 
normalization can be chosen as
\eqn\psnorm{
\langle (\l\g^m\t)(\l\g^n\t)(\l\g^p\t)(\t\g_{mnp}\t)\rangle = 1.
}
Although \psnorm\ involves only five $\t$'s it can be shown to
be supersymmetric \psf. Furthermore, given the fact that there is only one scalar in the
decomposition of $\l^3 \t^5$, any $\langle \l^\a\l^\b\l^\g f_{\a\b\g}(\t)\rangle$ can be determined
using symmetry arguments alone. This zero-mode integration has been automated in a {\tt FORM} \FORM\
program \PSS\ and therefore the component expansions of
any supersymmetric amplitude computed with the pure spinor formalism are readily obtained. However,
it is much more convenient to study the properties of the amplitudes directly in
the superspace expressions, where the BRST cohomology properties of the pure spinor superspace allow
several simplications to be carried out \refs{\mafraids,\FTAmps}.

\subsec{BRST building blocks}

In \WIP, the cohomology nature of pure spinor superspace has been exploited to define natural objects $T_i$, $T_{ij}$, $T_{ijk}$, $T_{ijkl}, \ldots$
which transform covariantly under the BRST charge of \BRSTc. These so-called 
BRST building blocks $T_{ijk{\ldots} }$ are constructed from the OPE's among the vertex operators
$V_i(z_i)U_j(z_j)U_k(z_k){\ldots}$ and are defined in such a way as to not contain any BRST-exact
terms \WIP. They are ultimately written in terms of super Yang-Mills superfields and therefore their
$\t$-expansions are also known; for example the building block
with one label is given by $T_i \equiv V_i$ where $V_i$ is the vertex operator of \vertices. The next 
building block $T_{ij}$ is an antisymmetric combination
\eqn\Td{
T_{ij} = {1\over 2}\(L_{ji} - L_{ij}\),
}
of the OPE residue $L_{ji} = - A^i_m (\l\g^m W^j) - V^i(k^i\cdot A^j)$ of $V_i(z_i)U_j(z_j)$. In \WIP, all 
building blocks up to and including $T_{ijklm}$ have been explicitly expressed in terms of SYM superfields. 
As already mentioned, they transform covariantly under the BRST charge \WIP,
\eqn\powerQTs{ \eqalign{
Q T_{12} &= s_{12} V_1 V_2, \ \ \ Q T_{123} = (s_{13}+s_{23}) T_{12} V_3 + s_{12}(V_1 T_{23} + T_{13} V_2)
\cr
QT_{12{\dots} n} &= \sum_{j=2}^{n} \sum_{\a \in P(\beta_j)} \( s_{1j} + s_{2j} + {\dots} + s_{j-1,j}\) 
T_{12{\ldots} j-1,\{\a\}}\; T_{j, \{\beta_j \backslash \a\}},
}}
where $\b_j = \{j+1,{\ldots},n\}$ and $P(\b_j)$ is the power set of $\b_j$. The $s_{ij} = k_i \cdot k_j$ denote standard Mandelstam variables.

Furthermore, the BRST building blocks $T_{i_1 i_2 \ldots i_p}$ have several symmetry properties which leave 
$(p-1)!$ independent permutations at rank $p$ and match the symmetries of the cubic diagrams they represent, e.g.
\eqn \Tsymm{
0 = T_{12} + T_{21} = T_{123} + T_{213} = T_{123} + T_{231} + T_{312}.
}
The rank $p$ object $T_{i_1 i_2 \ldots i_p}$ inherits all symmetries of its lower rank relative
$T_{i_1 i_2 \ldots i_{p-1}}$ in the first $p-1$ labels. It is therefore sufficient to give the novel relation 
at each rank which involves all the $p$ indices\foot{The notation $[i[jk]]$ using nested square brackets means 
consecutive antisymmetrization of pairs of labels starting from the outmost one, 
e.g. $T_{[i[jk]]} = 1/2(T_{i[jk]} - T_{[jk]i}) = 1/4(T_{ijk} - T_{ikj} - T_{jki} + T_{kji})$},
\eqn\TrankNu{\eqalign{
p=2n+1 &: \ \ \ T_{12\ldots n+1[n+2[\ldots [2n-1[2n,2n+1]] \ldots ]]} - 2 T_{2n+1\ldots n+2[n+1[\ldots [3[21]] \ldots ]]}  = 0 \cr
p=2n &: \ \ \ T_{12\ldots n[n+1[\ldots [2n-2[2n-1,2n]] \ldots ]]} + T_{2n\ldots n+1[n[\ldots [3[21]] \ldots ]]}  = 0,
}}
e.g. $T_{1234} + T_{1243} + T_{4321} + T_{4312} = 0$ at rank four.

Making use of these building blocks we have obtained the general form of the tree-level $n$ point correlator 
of \treepresc\ in \refs{\WIP,\MSSTFT} (see equation \stringBCJ\ below). As will now be discussed, the field-theory 
limit of the superstring amplitude can be used to find supersymmetric and local $n$-point BCJ-satisfying 
kinematic numerators \BCJ\ in a straightforward manner.

\newsec{String inspired BCJ numerators in superspace}

According to the hypothesis of BCJ, the color dressed $n$ point tree amplitude in gauge theories can be parametrized as
\eqn \organization{
{\cal A}_n = \sum_i \frac{c_i n_i}{\prod_{\alpha_i} s_{\alpha_i}} 
}
such that the kinematic factors $n_i$ satisfy Jacobi-like relations in one-to-one correspondence with the 
group-theoretic Jacobi identities for the color factors $c_i$,
\eqn\conjecDual{
c_i \pm c_j \pm c_k = 0 \quad \Rightarrow \quad n_i \pm n_j \pm n_k = 0.
}
The relative signs depend on the choice of signs when defining the color factors. The $i$ sum in \organization\ 
runs over the $(2n-5)!!$ cubic diagrams or pole channels specified by $n-3$ propagators $s_{\alpha_i}^{-1}$. 

In this section, we will derive the kinematic numerators $n_i$ in this organization scheme from superstring theory. 
Basic counting arguments together with relation between color ordered amplitudes imply that these $n_i$ satisfy the Jacobi identities \conjecDual\ by construction.

\topinsert
\centerline{
\tikzpicture[scale=0.9]
\draw node (0,0) {{ $ \displaystyle {\cal A}_4(1,2,3,4) \ \ =$}};
\scope[yshift=-0.5cm, xshift=0cm]
\draw [line width=0.30mm]  (2,0.5) -- (1.5,1) node[left]{$2$};
\draw [line width=0.30mm]  (2,0.5) -- (1.5,0) node[left]{$1$};
\draw [line width=0.30mm]  (2,0.5) -- (2.5,0.5) ;
\draw (2.25,0.7) node {$s$};
\draw [line width=0.30mm]  (2.5,0.5) -- (3,1) node[right]{$3$};
\draw [line width=0.30mm]  (2.5,0.5) -- (3,0) node[right]{$4$};
%
\endscope
\scope[xshift=-1cm]
\draw (4.5,0) node{$+$};
\draw [line width=0.30mm]  (5.5,0.25) -- (5,0.75) node[left]{$2$};
\draw [line width=0.30mm]  (5.5,-0.25) -- (5,-0.75) node[left]{$1$};
\draw [line width=0.30mm]  (5.5,0.25) -- (5.5,-0.25) ;
\draw (5.2,0) node {$u$};
\draw [line width=0.30mm]  (5.5,0.25) -- (6,0.75) node[right]{$3$};
\draw [line width=0.30mm]  (5.5,-0.25) -- (6,-0.75) node[right]{$4$};
%
\endscope
\scope [xshift = 7cm]
\draw node (0,0) {{ $ \displaystyle {\cal A}_4(1,3,2,4) \ \ =$}};
\scope[xshift=-3.5cm]
\draw [line width=0.30mm]  (5.5,-0.25) -- (5,0.75) node[left]{$2$};
\draw [line width=0.30mm]  (5.5,0.25) -- (5,-0.75) node[left]{$1$};
\draw [line width=0.30mm]  (5.5,0.25) -- (5.5,-0.25) ;
\draw (5.8,0) node {$t$};
\draw [line width=0.30mm]  (5.5,0.25) -- (6,0.75) node[right]{$3$};
\draw [line width=0.30mm]  (5.5,-0.25) -- (6,-0.75) node[right]{$4$};
%
\endscope
\scope[xshift=-1.5cm]
\draw (4.5,0) node{$-$};
\draw [line width=0.30mm]  (5.5,0.25) -- (5,0.75) node[left]{$2$};
\draw [line width=0.30mm]  (5.5,-0.25) -- (5,-0.75) node[left]{$1$};
\draw [line width=0.30mm]  (5.5,0.25) -- (5.5,-0.25) ;
\draw (5.2,0) node {$u$};
\draw [line width=0.30mm]  (5.5,0.25) -- (6,0.75) node[right]{$3$};
\draw [line width=0.30mm]  (5.5,-0.25) -- (6,-0.75) node[right]{$4$};
%
\endscope
\endscope
\endtikzpicture
}
\centerline{
\tikzpicture [scale=1.1]
\draw node (0,0) {{$ \displaystyle {\cal A}_5(1,2,3,4,5) \ \ =$}};
\scope[yshift=-0.5cm, xshift=0cm]
\draw [line width=0.30mm]  (2,0.5) -- (1.5,1) node[left]{$2$};
\draw [line width=0.30mm]  (2,0.5) -- (1.5,0) node[left]{$1$};
\draw [line width=0.30mm]  (2,0.5) -- (3,0.5) ;
\draw (2.25,0.3) node {$s_{12}$};
\draw [line width=0.30mm]  (2.5,0.5) -- (2.5,1) node[above]{$3$} ;
\draw (2.75,0.3) node {$s_{45}$};
\draw [line width=0.30mm]  (3,0.5) -- (3.5,1) node[right]{$4$};
\draw [line width=0.30mm]  (3,0.5) -- (3.5,0) node[right]{$5$};
%
\draw (4.9,0.5) node {$\displaystyle \quad + \quad{\rm cyclic}(12345)$};
\endscope
\endtikzpicture
} 
\tikzcaption\KKdiags{The four- and five-point amplitudes in terms of cubic graphs. 
The other five-point diagrams for the Kleiss-Kuijf 
basis can be obtained by relabeling.}
\endinsert

\subsec{The field theory setup}

The set of $(n-2)!$ color ordered amplitudes ${\cal A}_n(1,2_\rho,\ldots,(n-1)_\rho,n)$ with $\rho \in S_{n-2}$ and 
legs $n$ and $1$ fixed is sufficient to involve all of the $(2n-5)!!$ diagrams at least once. 
According to the Kleiss-Kuijf relations \KK\ all the other subamplitudes are sums over several ${\cal A}_n(1,2_\rho,\ldots,(n-1)_\rho,n)$
with coefficients $\pm 1$. That is why we will refer to them as the KK basis 
in the following. The $(n-2)!$ amplitudes at four-points are given by
$$
{\cal A}_4(1,2,3,4) = {n_s\over s} + {n_u\over u}, \qquad {\cal A}_4(1,3,2,4) = {n_t\over t} - {n_u\over u},
$$
while for five-points \BCJ,
\eqn\fiveptKK{
\eqalign{
{\cal A}_5(1,2,3,4,5) &= \frac{n_1}{s_{12}s_{45}} + \frac{n_2}{s_{23}s_{51}} + \frac{n_3}{s_{12}s_{34}} + \frac{n_4}{s_{23}s_{45}} + \frac{n_5}{s_{34}s_{51}}  \cr
{\cal A}_5(1,4,3,2,5) &= \frac{n_6}{s_{14}s_{25}} + \frac{n_5}{s_{34}s_{51}} + \frac{n_7}{s_{23}s_{14}} + \frac{n_8}{s_{34}s_{25}} + \frac{n_2}{s_{23}s_{51}}  \cr
{\cal A}_5(1,3,4,2,5) &= \frac{n_9}{s_{13}s_{25}} - \frac{n_5}{s_{34}s_{51}} + \frac{n_{10}}{s_{13}s_{24}} - \frac{n_8}{s_{34}s_{25}} + \frac{n_{11}}{s_{51}s_{24}}  \cr
{\cal A}_5(1,2,4,3,5) &= \frac{n_{12}}{s_{12}s_{35}} + \frac{n_{11}}{s_{24}s_{51}} - \frac{n_3}{s_{12}s_{34}} + \frac{n_{13}}{s_{35}s_{24}} - \frac{n_5}{s_{51}s_{34}}  \cr
{\cal A}_5(1,4,2,3,5) &= \frac{n_{14}}{s_{14}s_{35}} - \frac{n_{11}}{s_{24}s_{51}} - \frac{n_7}{s_{14}s_{23}} - \frac{n_{13}}{s_{24}s_{35}} - \frac{n_2}{s_{23}s_{51}}  \cr
{\cal A}_5(1,3,2,4,5) &= \frac{n_{15}}{s_{13}s_{45}} - \frac{n_2}{s_{23}s_{51}} - \frac{n_{10}}{s_{13}s_{24}} - \frac{n_4}{s_{23}s_{45}} - \frac{n_{11}}{s_{24}s_{51}} 
}}
and their representation in terms of cubic graphs is depicted in Fig.~1. One possible parametrization for the six point amplitude can be found in \MSST.

As mentioned in the introduction, the $n_i$ are not uniquely specified by the parametrization of ${\cal A}_4$ 
and ${\cal A}_5$ shown above. There still is a freedom to add zeros of the form $0= \left( {s_{ij} \over s_{ij}} - {s_{kl} \over s_{kl}} \right)\times (\ldots)$ 
to individual subamplitudes which amounts to reabsorbing contact terms into a different numerator $n_i$.

At four point level, the only Jacobi-like identity $n_s + n_t - n_u=0$ holds independent of the ambiguity of reshuffling 
contact terms between the numerators. This is a feature of the simple structure of ${\cal A}_4$ and its momentum phase space. 

In ${\cal A}_{5}$, imposing the color algebra on the fifteen $n_i$ yields nine independent relations
\eqn\Jacobifive{\eqalign{ 
0 &= n_3 - n_5 + n_8 \ = \ n_3 - n_1 + n_{12} \ = \ n_{10} - n_{11} + n_{13}  \cr
0 &= n_4 - n_2 + n_7 \ = \ n_4 - n_1 + n_{15} \ = \ n_{10} - n_9 + n_{15}  \cr
0 &= n_8 - n_6 + n_9 \ = \ n_5 - n_2 + n_{11} \ = \ n_7 - n_6 + n_{14} 
}}
which leave six independent numerators. However, the kinematic Jacobi identities at five and higher points generically 
fail to hold unless a specific class of choices for the $n_i$ is made.

 \centerline{
 \tikzpicture [scale=1.3]
 \scope[yshift=-0.5cm, xshift=-2.5cm]
 \draw [line width=0.30mm]  (2,0.5) -- (1.5,1) ;
 \draw (1.3,1.2) node {$a \ddots$};
 \draw [line width=0.30mm]  (2,0.5) -- (1.5,0) ;
 \draw (1.3,-0.1) node {$b \iddots$};
 \draw [line width=0.30mm]  (2,0.5) -- (2.5,0.5) ;
 \draw [line width=0.30mm]  (2.5,0.5) -- (3,1) ;
 \draw (3.2,1.2) node {$\iddots c$};
 \draw [line width=0.30mm]  (2.5,0.5) -- (3,0) ;
 \draw (3.2,-0.1) node {$\ddots d$};
 \draw (2.25,-0.5) node {$\displaystyle  \frac{c_i n_i}{s_i}$}; 
 \endscope
 \scope[xshift=-2.5cm]
 \draw (4,0) node{$+$};
 \draw [line width=0.30mm]  (5.5,-0.25) -- (5,0.75) ;
 \draw (4.8,0.95) node {$b \ddots$};
 \draw [line width=0.30mm]  (5.5,0.25) -- (5,-0.75) ;
 \draw (4.8,-0.85) node {$a \iddots$};
 \draw [line width=0.30mm]  (5.5,0.25) -- (5.5,-0.25) ;
 \draw [line width=0.30mm]  (5.5,0.25) -- (6,0.75) ;
 \draw (6.2,0.95) node {$\iddots c$};
 \draw [line width=0.30mm]  (5.5,-0.25) -- (6,-0.75);
 \draw (6.2,-0.85) node {$\ddots d$};
 \draw (5.5,-1) node {$\displaystyle  \frac{c_j n_j}{s_j}$}; 
 \endscope
 \scope[xshift=0.5cm]
 \draw (4,0) node{$-$};
 \draw [line width=0.30mm]  (5.5,0.25) -- (5,0.75) ;
 \draw (4.8,0.95) node {$b \ddots $};
 \draw [line width=0.30mm]  (5.5,-0.25) -- (5,-0.75) ;
 \draw (4.8,-0.85) node {$a\iddots $};
 \draw [line width=0.30mm]  (5.5,0.25) -- (5.5,-0.25) ;
 \draw [line width=0.30mm]  (5.5,0.25) -- (6,0.75) ;
 \draw (6.2,0.95) node {$\iddots c$};
 \draw [line width=0.30mm]  (5.5,-0.25) -- (6,-0.75);
 \draw (6.2,-0.85) node {$\ddots d$};
 \draw (5.5,-1) node {$\displaystyle  \frac{c_k n_k}{s_k}$}; 
 \endscope
 \endtikzpicture
 }
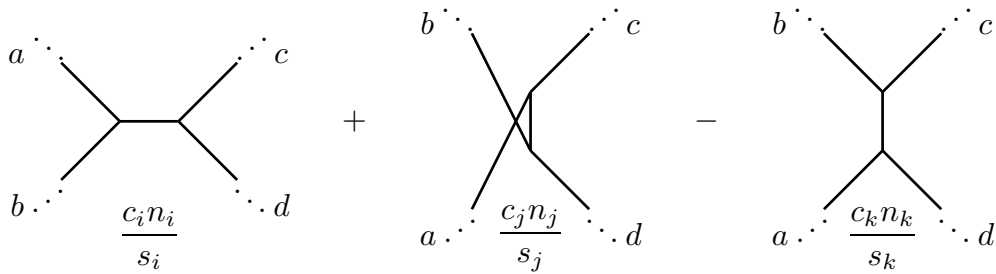
\tikzcaption \figtriplet{Triplet of subdiagrams where the sum over the associated color factor vanishes due to the Jacobi identity $f^{e[ab} f^{c]de}=0$.}

More generally, if a suitable parametrization and contact term bookkeeping is chosen the duality between 
color and kinematics manifests itself for each triplet of subdiagrams shown in figure \figtriplet. For cubic 
graphs describing an $n$-point tree amplitude, there can be arbitrary further subdiagrams $a,b,c,d$ attached to 
the dotted lines with $n-4$ common propagators in total.

In an $n$-point tree level amplitude, if all the Jacobi-like identities for the $(2n-5)!!$ numerators in 
${\cal A}_n$ are satisfied, then a set of $(n-2)!$ independent $n_i$ remains. In the following we will reverse 
the line of reasoning: If one can show that the KK basis ${\cal A}_n(1,2_\s,\ldots, (n-1)_\s,n)$ of color ordered 
amplitudes can be expressed in terms of $(n-2)!$ basis numerators, then there must be as many equations between the 
larger set of $(2n-5)!!$ numerators as there are Jacobi identities.

\subsec{A minimal kinematic basis from superstring theory}

Supersymmetric field theory tree amplitudes can also be obtained from the low--energy limit of 
superstring theory where the dimensionless combinations $\alpha' s_{i_1 \ldots  i_p}$ of Regge 
slope~$\alpha'$ and Mandelstam bilinears $s_{i_1 \ldots  i_p} = {1\over 2} (k_{i_1} + k_{i_2} + \ldots + k_{i_p})^2$ 
are formally sent to zero. Using the pure spinor formalism \psf, we will show in \WIP\ that the color stripped 
superstring $n-$point 
amplitude is given by
\eqn\stringBCJ{
\eqalign{
&{\cal A}_n^{{\rm string}}(1_\s,2_\s,\ldots,(n-1)_\s,n;\alpha') =(2\ap)^{n-3} \prod_{i=2}^{n-2} 
\int_{{\cal I}_\s} \! \!  {\rm d} z_i \ \prod_{j<k} |z_{jk}|^{-2 \alpha' s_{jk}}    \cr
& \ \ \ \ \sum_{j=1}^{n-2} {\langle T_{12\ldots j}  \, T_{n-1,n-2 \ldots j+1} \, V_n 
\rangle \over (z_{12} z_{23} \ldots  z_{p-1,p}) (z_{n-1,n-2}  z_{n-2,n-3} \ldots  z_{j+2,j+1})}
+ {\cal P}(2,3,\ldots ,n-2) 
}}
in terms of the BRST building blocks $T_{12\ldots p}$, with 
$z_{jk}=z_j-z_k$. The ${\cal P}(2,3,\ldots ,n-2)$ denotes the sum over all permutations of the labels $2,3,\ldots,n-2$ of the integrated vertex operators \vertices. 

The ordering $\s \in S_{n-1}$ of the external legs is reflected in the 
integration region ${\cal I}_\s$ for the worldsheet positions $z_{2_\s},z_{3_\s},\ldots,z_{(n-2)_\s}$, 
the remaining ones are fixed as $(z_{1_\s},z_{(n-1)_\s},z_n) = (0,1,\infty)$ by $SL(2,R)$ invariance of 
the disk worldsheet. More precisely, ${\cal I}_{\s}$ is defined such that only those $z_i$ which respect 
the ordering
$$
0=z_{1_\s} \leq z_{2_\s} \leq z_{3_\s} \leq \ldots \leq z_{(n-2)_\s} \leq z_{(n-1)_\s} = 1
$$
are integrated over.

To arrive at \stringBCJ\ from the pure spinor conformal field theory (CFT), one has to reexpress integrals with double pole factors 
like $z_{ij}^{-2}$ in terms of single pole integrals (with integrands such as $(z_{ij} z_{jk})^{-1}$). The superfields 
associated with these world-sheet poles conspire such that the OPE residues of $V_1(z_1) U_2(z_2) \ldots U_p(z_p)$ in
single pole integrals receive all the corrections necessary to form BRST building blocks $T_{12\ldots p}$. It has 
already been realized in \refs{\FivePt,\ExtendedBCJs} that double poles cause various technical complications and 
in particular prevent the basis of kinematics to boil down to the desirable size of $(n-2)!$.

The $\alpha' \rightarrow 0$ limit of \stringBCJ\ extracts propagators of cubic field theory diagrams from the $n-3$ 
fold worldsheet integrals. Adjusting the integration region ${\cal I}_\s$ to the color ordering makes sure
that the integrals in ${\cal A}_n^{{\rm string}}(1_\s,2_\s,\ldots,(n-1)_\s,n;\alpha')$ only generate those
pole channels which appear in the corresponding field theory amplitude ${\cal A}_n(1_\s,2_\s,\ldots,(n-1)_\s,n)$.
A method  to  efficiently extract the field theory limit of a general $n$-point integral is described in \WIP. 

The remarkable property of \stringBCJ\ in view of the BCJ organization is the number of independent superfield
kinematics $\langle T_{12\ldots j}  T_{n-1,n-2,\ldots,j+1} V_n \rangle$. Each of the $n-2$ terms in the $j$ sum of
\stringBCJ\ involves $(n-3)!$ permutations of $\langle T_{12\ldots j}  T_{n-1,n-2,\ldots,j+1} V_n \rangle$ in
the legs $(2,3,\ldots,n-2)$ such that we have $(n-2)!$ kinematic packages in total. The world-sheet integrand remains
the same for any color ordering, only the integration region ${\cal I}_\s$ changes between the subamplitudes. 

Hence, the $(n-2)!$ basic kinematics $\langle T_{12\ldots j}  T_{n-1,n-2,\ldots,j+1} V_n \rangle$ combine to the 
kinematic factors for any color stripped superstring amplitude, and in particular, they generate all the $(2n-5)!!$ 
BCJ numerators of the field theory amplitude in the $\ap \rightarrow 0$ limit. Their coefficients are determined by the
pole structure of the integrals in the corresponding integration region ${\cal I}_\s$ which is specified by the color 
ordering of ${\cal A}_n(1,2_\s,3_\s,\ldots,(n-1)_\s,n)$. 

As we have argued in the previous subsection -- having a set of no more than $(n-2)!$ independent numerators is necessary 
for imposing the Jacobi-like identities (dual to color factors) on the $(2n-5)!!$ numerators of the pole channels in 
various subamplitudes. In the next paragraph we explain why the number $(n-2)!$ of kinematics in \stringBCJ\ is 
also sufficient to satisfy the Jacobi relations.

\subsec{The vanishing of numerator triplets}

The fact that only $(n-2)!$ BCJ numerators can be linearly independent implies the existence of as many linear 
homogeneous relations between the $n_i$ as there are Jacobi identities. Since the field theory limits of the 
integrals in \stringBCJ\ involve no other coefficients than $0$ and $\pm 1$ for the propagators, these relations must be of the form
\eqn \enoughrel{
n_{i_1} \pm n_{i_2} \mp n_{i_3} \pm \ldots \mp n_{i_{p-1}} \pm n_{i_p} = 0,
}
with a so far unspecified number $p$ of terms. In order to show that they can always be arranged into vanishing 
statements for triplets $n_{i_1} \pm n_{i_2} \mp n_{i_3} = 0$ one has to make use of the monodromy relations of field theory \BCJ
\eqn \monodromy{
s_{12} {\cal A}_n(2,1,3,\ldots ,n) + \sum_{j=3}^{n-1} \left( \sum_{k=2}^{j} s_{1k} \right) {\cal A}_n(2,3,\ldots,j,1,j+1,\ldots,n-1 ,n) = 0,
}
which allow to reduce the KK subamplitudes to a basis of $(n-3)!$ independent ones. Their string theory generalization
replaces the sum $\sum_k s_{1k}$ by 
sine functions $\sin \left( 2\alpha' \pi \sum_k s_{1k} \right)$ \refs{\monodVanhove,\monodStie}.

By taking appropriate permutations of \monodromy\ and decomposing the occurring subamplitudes in pole channels, one 
can derive identities between Jacobi triplets $(n_{i_k} , n_{i_l} ,n_{i_m})$ dual to color 
factors with $c_{i_k} + c_{i_l} + c_{i_m} = 0$ of the following form \refs{\ExtendedBCJs,\tye}
\eqn \monodr{
\sum_{i} { n_{i_k} + n_{i_l} + n_{i_m} \over \prod_{\alpha_i}^{n-4} s_{\alpha_i}} = 0.
}
The $i$ sum runs over $n-1$ point channels of total number $2^{n-3} (2n-7)!! (n-3)/ (n-2)!$ and involves the $n-4$ 
propagators $s_{\alpha_i}$ common among the $n_{i_k}$, $n_{i_l}$ and $n_{i_m}$ channels. As a consquence, $n_{i_k} + n_{i_l} + n_{i_m}$ 
must vanish at the residue of the $n-4$ poles, independent on the assignment of contact terms to the numerators.

Suppose the linear dependences \enoughrel\ failed to make Jacobi triplets of BCJ numerators vanish, i.e. $p>3$, then \monodr\ 
would involve terms 
$$
{ n_{i_1} + n_{i_2} + n_{i_3} \over \prod_{\alpha_i}^{n-4} s_{\alpha_i}} = - { \sum_{j=4}^{p} n_{i_j} \over \prod_{\alpha_i}^{n-4} s_{\alpha_i}}
$$
where each noncancelling $n_{i_{j>3}}$ is muliplied with at least one propagator outside its channels. This is clearly 
incompatible with \monodr\ because there will remain contributions with a specific set of $n-4$ poles 
from each term like that\foot{The claimed incompatibility rests on the linear independence of the $(n-2)!$ factorial 
basis numerators. We wish to thank Henrik Johansson for pointing out that a loophole would arise otherwise.}.

The conclusion is that the $(2n-5)!! - (n-2)!$ relations \enoughrel\ can always be brought into a form that reproduces 
all the dual Jacobi identities for the kinematic numerators $n_i$. If this was not the case, inconsistencies would arise 
in the monodromy relations \monodromy\ or \monodr\ between color ordered field theory amplitudes. Hence, the number $(n-2)!$ 
of kinematics in \stringBCJ\ guarantees that all the $n_i$ constructed from the $\ap \rightarrow 0$ limit of the integrals 
over ${\cal I}_\s$ satisfy the Jacobi-like relations $n_i+n_j+n_k=0$.

\subsec{The explicit formula for BCJ numerators}
\subseclab\KsSec\

The worldsheet integrand of \stringBCJ\ suggests to label the $(n-2)!$ basis kinematics in an $n$-point amplitude 
by an $S_{n-3}$ permutation $\s$ and an integer $l=1,\ldots,n-2$:
\eqn\Kbasic{
{\cal K}^l_\s = \langle T_{12_\s 3_\s \ldots l_\s} T_{n-1,(n-2)_\s \ldots (l+1)_\s} V_n \rangle , \ \ \ l=1,\ldots,n-2, \ \ \ \s \in S_{n-3}
}
This makes sure that the residual $S_{n-3}$ relabelling symmetry stays visible in the KK basis of the field 
theory limit. As we have mentioned before, knowing all the KK subamplitudes ${\cal A}_n(1,2_\rho,\ldots,(n-1)_\rho,n)$ is 
sufficient to address each channel and to thereby identify all the $(2n-5)!!$ numerators $n_i$. The superstring amplitude 
\stringBCJ\ provides a general prescription to construct these KK subamplitudes in terms of the basis kinematics 
${\cal K}^l_\s$ defined by \Kbasic\
$$
{\cal A}_n(1,2_\rho,\ldots,(n-1)_\rho,n)
= \sum_{l=1}^{n-2} \sum_{\s \in S_{n-3}} P_\rho{}^{(l,\s)} \, {\cal K}^l_\s
$$
This introduces a $(n-2)! \times (n-2)!$ matrix $P_\rho{}^{(l,\s)}$ of kinematic poles whose entries is determined by the integral of the 
worldsheet polynomial associated with $(l,\s)$ over the integration region ${\cal I}_\rho$:
\eqn\defP{
P_\rho{}^{(l,\s)} = \lim_{\ap \rightarrow 0} (2\ap)^{n-3} \prod_{i=2}^{n-2} \int_{{\cal I}_\rho} \! \!  {\rm d} z_i \ 
\frac{\prod_{j<k} |z_{jk}|^{-2 \alpha' s_{jk}} }{z_{12_\s} z_{2_\s 3_\s} \ldots z_{(l-1)_\s l_\s} \ z_{n-1,(n-2)_\s} \ldots z_{(l+2)_\s,(l+1)_\s}}
}
These $\ap \rightarrow 0$ limits can be straightforwardly evaluated using the methods of \WIP.

The idea of introducing an $(n-2)!$ vector of KK amplitudes and relating it to $(n-2)!$ independent numerators by a square 
matrix already appeared in \VamanP. In our situation, the basis \Kbasic\ of numerators is set by the superstring 
computation, and our $P_\rho{}^{(l,\s)}$ matrix is a specialization of the propagator matrix $M$ in this reference to 
the pure spinor basis of kinematics. The linear dependences of KK subamplitudes due to BCJ relations \monodromy\ imply that the 
$(n-2)! \times (n-2)!$ propagator matrices $M$ or $P_\rho{}^{(l,\s)}$ have rank $(n-3)!$.
 
Not all the entries of the pole matrix $P_\rho{}^{(l,\s)}$ have to be computed separately. The following trick 
relates many of them by relabelling and thereby reduces the computational effort on the way towards explicit BCJ numerators: Exclude 
the leg $n-1$ from the $\rho \in S_{n-2}$ permutations specifying the KK subamplitudes. They then fall into $n-2$ 
classes ${\cal A}_n (1,2_\s , \ldots, j_\s,n-1,(j+1)_\s,\ldots,(n-2)_\s, n)$ with $j-1$ legs between $1$ and 
$n-1$ and another $n-2-j$ legs between $n-1$ and $n$. The legs $2,3,\ldots,n-2$ are interchanged by $S_{n-3}$ 
permutations $\s$ like in \fiveptmintwo. It is sufficient to compute one representative of the $n-2$ classes of
KK amplitudes, the others then follow as $S_{n-3}$ permutations in $2,3,\ldots,n-2$. More precisely, 
once the first $n-2$ columns of \defP\ with $\rho =(2,3,\ldots,j,n-1,j+1,\ldots,n-2)$ and $j=1,2,\ldots,n-2$ are 
known, then the others follow from
\eqn\Prel{
P_{\rho= (2_\s,3_\s,\ldots,j_\s,n-1,(j+1)_\s,\ldots,(n-2)_\s)}{}^{(l,\tau)} = P_{\rho = (2,3,\ldots,j,n-1,j+1 ,
\ldots,n-2) }{}^{(l,\s^{-1} \circ \tau)} \, \Bigl. \Bigr|_{k_i \mapsto k_{\s(i)}}
}
where the concatenation of $\s^{-1},\tau \in S_{n-3}$ is to be understood as $(\s^{-1} \circ \tau)(i) = \s^{-1}( \tau(i))$. 
The proof of \Prel\ is a simple matter of bookkeeping with world-sheet integration variables in \defP.

This relabelling strategy reduces the number of independent evaluations of \defP\ from $(n-2)! \times (n-2)!$ down to 
$(n-2) \times (n-2)!$, i.e. the work at this step is reduced by a factor of $(n-3)!$. But the success of these $S_{n-3}$ 
relabellings does not extend to the leg $n-1$. The $P_\rho{}^{(l,\s)}$ entries for the $n-2$ classes of KK 
subamplitudes ${\cal A}_n (1,2_\s , \ldots, j_\s,n-1,(j+1)_\s,\ldots,(n-2)_\s, n)$ with $j=1,2,\ldots,n-2$ have 
a different number and structure of terms as $j$ varies. This will become more obvious from the examples in the 
next section. As a consequence, the $n_i$ appearing in these subamplitudes involve more basic kinematics ${\cal K}^l_{\s}$ 
and do not follow from other BCJ numerators by relabelling. 


\subsec{Jacobi-friendly notation}

At higher points it is not convenient to denote the BCJ numerators sequentially by $n_{i}$ for $i=1,2,\ldots (2n-5)!!$. 
Already the presentation of the fifteen numerators in the five point KK basis \fiveptKK\ suffers from the arbitrary assignment 
of numbers 1 to 15 to the cubic diagrams. It is not at all obvious from their labels which of them combine to form the Jacobi
triplets \Jacobifive. 

Instead, we will use a notation introduced by \VamanP\ which reflects the structure of the diagram and allows
the associated propagators to be reconstructed. More importantly, it makes the dual structure constant contraction 
available from which one can infer the symmetry properties $f^{abc} = - f^{bac}$ and the Jacobi identities 
$f^{b[a_1 a_2} f^{a_3] bc} = 0$. Let us clarify these properties by explicit examples:

The four points amplitude encompasses three diagrams of the form

\centerline{
\tikzpicture[scale=1]
\scope[yshift=-0.5cm, xshift=0cm]
\draw [line width=0.30mm]  (2,0.5) -- (1.5,1) node[left]{$2$};
\draw [line width=0.30mm]  (2,0.5) -- (1.5,0) node[left]{$1$};
\draw [line width=0.30mm]  (2,0.5) -- (2.5,0.5) ;
\draw [line width=0.30mm]  (2.5,0.5) -- (3,1) node[right]{$3$};
\draw [line width=0.30mm]  (2.5,0.5) -- (3,0) node[right]{$4$};
\endscope
\draw (6.5,0) node{$\sim \displaystyle {1 \over s_{12}} (f^{a_1 a_2 b} f^{b a_3 a_4} ) \times n[12,34] = { c_s n_s \over s} $};
\endtikzpicture
}
\noindent
where $n[12,34]$ has the same symmetries as the structure constants involved:
$$
n[ij,kl] =-n[ji,kl]  = -n[ij,lk] = n[ji,lk] , \ \ \ n[ij,kl] = n[kl,ij]   
$$
If we assign $n_t = n[13,42]$ and $n_u = n[23,41]$, then the Jacobi identity $n_s+ n_t - n_u = 0$ can 
be written more compactly as
$$
n[1 \{ 2,34\}] := n[12,34]+n[13,42]+n[14,23]=0 .
$$
At five points, the first out of fifteen pole channels contributes

\centerline{
\tikzpicture [scale=1.1]
\scope[yshift=-0.5cm, xshift=0cm]
\draw [line width=0.30mm]  (2,0.5) -- (1.5,1) node[left]{$2$};
\draw [line width=0.30mm]  (2,0.5) -- (1.5,0) node[left]{$1$};
\draw [line width=0.30mm]  (2,0.5) -- (3,0.5) ;
\draw [line width=0.30mm]  (2.5,0.5) -- (2.5,1) node[above]{$3$} ;
\draw [line width=0.30mm]  (3,0.5) -- (3.5,1) node[right]{$4$};
\draw [line width=0.30mm]  (3,0.5) -- (3.5,0) node[right]{$5$};
\draw (7.7,0.5) node {$\displaystyle \sim {1 \over s_{12} s_{45}}  (f^{a_1 a_2 b} f^{ba_3c} f^{c a_4 a_5} ) \times n[12,3,45] = \frac{c_1 n_1}{s_{12} s_{45}} $};
\endscope
\endtikzpicture
}
\noindent
where the kinematic numerators inherit their antisymmetry under flipping a cubic vertex from the structure constants:
$$
n[ij,k,lm] =-n[ji,k,lm]  = -n[ij,k,ml] = n[ji,k,ml] , \ \ \ n[ij,k,lm] = -n[lm,k,ij]   
$$
All the Jacobi identities \Jacobifive\ can be diagrammatically found by attaching a cubic vertex with two external 
legs to one of the dotted lines of figure \figtriplet. They can be cast into unified form
$$
n[ij,\{ k, lm\}] = 0.
$$

Six point amplitudes introduce two topologies of cubic diagrams

\centerline{
\tikzpicture [scale=1.1]
\scope[yshift=-0.5cm, xshift=0cm]
\draw [line width=0.30mm]  (2,0.5) -- (1.5,1) node[left]{$2$};
\draw [line width=0.30mm]  (2,0.5) -- (1.5,0) node[left]{$1$};
\draw [line width=0.30mm]  (2,0.5) -- (3.5,0.5) ;
\draw [line width=0.30mm]  (2.5,0.5) -- (2.5,1) node[above]{$3$} ;
\draw [line width=0.30mm]  (3,0.5) -- (3,1) node[above]{$4$} ;
\draw [line width=0.30mm]  (3.5,0.5) -- (4,1) node[right]{$5$};
\draw [line width=0.30mm]  (3.5,0.5) -- (4,0) node[right]{$6$};
\draw (8.3,0.5) node {$\displaystyle \sim {1 \over s_{12} s_{123} s_{56}}  (f^{a_1 a_2 b} f^{b a_3 c} f^{c a_4 d} f^{d a_5 a_6 }  ) \times n[12,3,4,56] $};
\endscope
\endtikzpicture
}

\centerline{
\tikzpicture [scale=1.0]
\scope[yshift=-0.5cm, xshift=0cm]
\draw [line width=0.30mm]  (2,0.5) -- (1.5,1) node[left]{$2$};
\draw [line width=0.30mm]  (2,0.5) -- (1.5,0) node[left]{$1$};
\draw [line width=0.30mm]  (2,0.5) -- (3,0.5) ;
\draw [line width=0.30mm]  (2.5,0.5) -- (2.5,1) ;
\draw [line width=0.30mm]  (2.5,1) -- (2,1.5) node[left]{$3$} ;
\draw [line width=0.30mm]  (2.5,1) -- (3,1.5) node[right]{$4$} ;
\draw [line width=0.30mm]  (3,0.5) -- (3.5,1) node[right]{$5$};
\draw [line width=0.30mm]  (3,0.5) -- (3.5,0) node[right]{$6$};
\draw (8.2,0.5) node {$\displaystyle \sim {1 \over s_{12} s_{34} s_{56}}   (f^{a_1 a_2 b} f^{a_3 a_4 c} f^{a_5 a_6 d} f^{bcd} ) \times n[12,34,56] $};
\endscope
\endtikzpicture
}

\noindent
which imprint the following symmetries on the BCJ numerators:
\eqn \sixsymm{ \eqalign{
n[ij,k,l,mp] &=-n[ji,k,l,mp]  , \ \ \ n[ij,k,l,mp] = n[mp,l,k,ij]   \cr
n[ij,kl,mp] &=-n[ji,kl,mp]  , \ \ \ n[ij,kl,mp] = -n[kl,ij,mp]
}}
Also the Jacobi identites exhibit different topologies here, one can either attach three 
point vertices to two different external lines of \figtriplet\ or one color ordered four point 
diagram to one external line:
$$
n[ij,k, \{ l,mp \}] = 0 , \ \ \ n[ij,kl,mp] = n[ij,k,l,mp] -n[ij,l,k,mp] 
$$
The latter expresses any diagram of snowflake shape in terms of the other topology.

Seven points again introduce two topologies of cubic diagrams

\centerline{
\tikzpicture [scale=1.1]
\scope[yshift=-0.5cm, xshift=0cm]
\draw [line width=0.30mm]  (2,0.5) -- (1.5,1) node[left]{$2$};
\draw [line width=0.30mm]  (2,0.5) -- (1.5,0) node[left]{$1$};
\draw [line width=0.30mm]  (2,0.5) -- (4,0.5) ;
\draw [line width=0.30mm]  (2.5,0.5) -- (2.5,1) node[above]{$3$} ;
\draw [line width=0.30mm]  (3,0.5) -- (3,1) node[above]{$4$} ;
\draw [line width=0.30mm]  (3.5,0.5) -- (3.5,1) node[above]{$5$} ;
\draw [line width=0.30mm]  (4,0.5) -- (4.5,1) node[right]{$6$};
\draw [line width=0.30mm]  (4,0.5) -- (4.5,0) node[right]{$7$};
\draw (8.3,0.5) node {$\displaystyle \sim {f^{a_1 a_2 b} f^{b a_3 c} f^{c a_4 d} f^{d a_5 e } f^{e a_6 a_7} \over s_{12} s_{123} s_{567} s_{67}}  n[12,3,4,5,67] $};
\endscope
\endtikzpicture
}

\centerline{
\tikzpicture [scale=1.0]
\scope[yshift=-0.5cm, xshift=0cm]
\draw [line width=0.30mm]  (2,0.5) -- (1.5,1) node[left]{$2$};
\draw [line width=0.30mm]  (2,0.5) -- (1.5,0) node[left]{$1$};
\draw [line width=0.30mm]  (2,0.5) -- (3,0.5) ;
\draw [line width=0.30mm]  (2.5,0.5) -- (2.5,1) node[above]{$3$} ;
\draw [line width=0.30mm]  (3,0.5) -- (3.5,1) ;
\draw [line width=0.30mm]  (3,0.5) -- (3.5,0) ;
\draw [line width=0.30mm]  (3.5,1) -- (4,1.5) node[right]{$4$};
\draw [line width=0.30mm]  (3.5,1) -- (4,1) node[right]{$5$};
\draw [line width=0.30mm]  (3.5,0) -- (4,0) node[right]{$6$};
\draw [line width=0.30mm]  (3.5,0) -- (4,-0.5) node[right]{$7$};
\draw (8,0.5) node {$\displaystyle \sim {f^{a_1 a_2 b} f^{b a_3  c} f^{cde} f^{d a_4 a_5} f^{e a_6 a_7} \over s_{12}s_{123} s_{45} s_{67}}   
n \left[ 12,3, 45, 67  \right] $};
\endscope
\endtikzpicture
}

\noindent
They give rise to symmetry properties
\eqn \sevensymm{ \eqalign{
n[ij,k,l,m,pq] &=-n[ji,k,l,m,pq]  , \ \ \ n[ij,k,l,m,pq] = - n[pq,m,l,k,ij] 
\cr
n \left[ ij,k, lm, pq  \right] = - n& \left[ ji,k, lm, pq  \right] 
= - n \left[ ij,k, ml, pq  \right]   , \ \ \ n \left[ ij,k, lm, pq  \right]  
= - n \left[ ij,k, pq,lm  \right] 
}}
and Jacobi identities which eliminate the topology with more branchings:
$$
n[ij,k,l,\{m,pq\}]= 0, \ \ \ n \left[ ij,k, lm, pq  \right] = n[ij,k,l,m,pq] - n[ij,k,m,l,pq]
$$
The four different topologies at eight points can still be captured by the suggestive notations 
$n[ij,k,l,m,p,qr],n[ij,k,lm,p,qr], n \left[ ij,k, l, mp, qr  \right]$ and 
$n \left[  kl,  ij , mp, qr  \right]$. Jacobi identities relate diagrams of 
different topology such that all of them can be represented in terms of the simplest numerators $n[ij,k,l,m,p,qr]$.

\newsec{Explicit examples}

In order to make the very general statements more tractable, we shall analyze explicit examples up to seven-point in detail. 

\subsec{Four-point}

Let us first 
of all compute the four-point numerators as a warm-up exercise. For $n=4$, the general formula \stringBCJ\ gives rise to one 
worldsheet integral and two $z_2$ functions in the integrand,
$$
{\cal A}_4^{{\rm string}}(1,2_\rho,3_\rho,4;\alpha') = 2\alpha' \int_{{\cal I}_\rho} {\rm d} z_2 \, 
|z_2|^{-2\alpha's}  |1-z_2|^{-2\alpha' u} \left\{ {\langle T_{12} V_3 V_4 \rangle \over z_2} + { \langle V_1 T_{23} V_4 \rangle \over 1-z_2} \right\}
$$
which can be easily evaluated in terms of the Euler Beta function. We obtain the following entries for the propagator 
matrix \defP\ by taking the field theory limit:
\eqn \fourptpole{ \eqalign{
(P_{(2,3)}{}^{1},P_{(2,3)}{}^{2}) &= \lim_{\alpha' \rightarrow 0} 2\alpha' \int_{0}^1 {\rm d} z_2 \, |z_2|^{-2\alpha's}  |1-z_2|^{-2\alpha' u}  
\left( {1 \over z_2} , {1 \over 1-z_2} \right) = \left( {1 \over s} , {1 \over u} \right)
 \cr
 (P_{(3,2)}{}^{1},P_{(3,2)}{}^{2}) &= \lim_{\alpha' \rightarrow 0} 2\alpha' \int_{1}^{\infty} \! \! {\rm d} z_2 \, |z_2|^{-2\alpha's}  |1-z_2|^{-2\alpha' u}  
 \left( {1 \over z_2} , {1 \over 1-z_2} \right) = \left( {1 \over t} , -{1 \over u}-{1 \over t} \right)
 }}
The $S_{n-3}$ permutation $\s$ which appears as a superscript label of the general $P_\rho{}^{(l,\s)}$ becomes trivial 
at four points. This leads to field theory subamplitudes
$$
{\cal A}_4 (1,2,3,4) = {n_s \over s} + {n_u \over u} , \ \ \ {\cal A}_4 (1,3,2,4) = {n_s - n_u \over t} - {n_u \over u} = -{n_t \over t} - {n_u \over u}
$$
with BCJ numerators that manifestly satisfy a Jacobi relation:
$$
n_s = \langle T_{12} V_3 V_4 \rangle, \ \ \ n_u = \langle V_1 T_{23} V_4 \rangle, \ \ \ n_t =n_u - n_s =\langle V_1 T_{23} V_4 \rangle - \langle T_{12} V_3 V_4 \rangle 
$$
They are evaluated in superfield components in appendix C.

\subsec{Five-point}

The color ordered five-point superstring amplitude encompasses six basis kinematics:
\eqn\stringfivept{
{\cal A}_5^{{\rm string}}(1,2_\rho,3_\rho,4_\rho,5;\alpha') =(2\ap)^{2} \int_{{\cal I}_\rho} \! \!  {\rm d} z_2 \, {\rm d} z_3 \, \prod_{j<k} |z_{jk}|^{-2 \alpha' s_{jk}}
}
$$
 \left( \frac{ \langle T_{123} V_4 V_5 \rangle }{z_{12} z_{23}} + \frac{ \langle T_{132} V_4 V_5 \rangle }{z_{13} z_{32}} 
 + \frac{ \langle T_{12} T_{43} V_5 \rangle }{z_{12} z_{43}}+ \frac{ \langle T_{13} T_{42} V_5 \rangle }{z_{13} z_{42}}
 + \frac{ \langle V_{1} T_{432} V_5 \rangle }{z_{43} z_{32}}+ \frac{ \langle V_{1} T_{423} V_5 \rangle }{z_{42} z_{23}} \right)
$$
The double pole $z_{23}^{-2}$ appearing in the five point integrand of other references \refs{\FivePt,\ExtendedBCJs} was absorbed
into the single-pole terms such that BRST building blocks $T_{ij}$ and $T_{ijk}$ could be formed. Absence of double 
poles is crucial for satisfying all the dual Jacobi relations and its removal was essentially dictated by the
BRST cohomology properties of those building blocks \refs{\WIP,\MSST}.

Let us rewrite the superfields in the ${\cal K}^l_\s$ notation of subsection \KsSec\ in order to make better use of the residual 
relabelling symmetry in $2 \leftrightarrow 3$:
\eqn\shortfivept{\eqalign{
{\cal K}^3_{(23)} = & \langle T_{123} V_4 V_5 \rangle , \ \ \ {\cal K}^2_{(23)} \ = 
\ \langle T_{12} T_{43} V_5 \rangle , \ \ \ {\cal K}^1_{(23)} \ = \ \langle V_{1} T_{432} V_5 \rangle \cr
{\cal K}^3_{(32)} = & \langle T_{132} V_4 V_5 \rangle , \ \ \ {\cal K}^2_{(32)} \ = 
\ \langle T_{13} T_{42} V_5 \rangle , \ \ \ {\cal K}^1_{(32)} \ = \ \langle V_{1} T_{423} V_5 \rangle
}}
Performing the field theory limit of the integrals \stringfivept\ gives rise to the following six KK subamplitudes 
(where the permutation $\s$ of $2$ and $3$ can be kept general because of \Prel)
\eqn\fiveptmintwo{
\eqalign{
{\cal A}_5(1,2_\s,3_\s,4,5) =& 
  {{\cal K}^{3}_{\s(23)} \over s_{45} s_{12_\s}}
+ {{\cal K}^1_{\s(23)} - {\cal K}^1_{\s(32)}  \over s_{51} s_{2_\s3_\s}}
- { {\cal K}^2_{\s(23)} \over s_{12_\s} s_{3_\s 4}}
+ {  {\cal K}^3_{\s(23)} - {\cal K}^3_{\s(32)} \over s_{2_\s 3_\s} s_{45}}
+ { {\cal K}^1_{\s(23)} \over s_{3_\s 4} s_{51}} \cr
{\cal A}_5(1,2_\s,4,3_\s,5) =& 
{ {\cal K}^3_{\s(23)} + {\cal K}^2_{\s(23)} \over s_{12_\s} s_{3_\s 5}} 
- {  {\cal K}^1_{\s(32)} \over s_{2_\s 4} s_{51}} 
+ {{\cal K}^2_{\s(23)}  \over s_{3_\s 4} s_{12}} 
- {  {\cal K}^2_{\s(32)} + {\cal K}^1_{\s(32)} \over s_{3_\s 5} s_{2_\s 4}} 
- { {\cal K}^1_{\s(23)} \over s_{51} s_{3_\s 4}} \cr
{\cal A}_5(1,4,2_\s,3_\s,5) =& 
  {{\cal K}^3_{\s(23)} + {\cal K}^2_{\s(23)} + {\cal K}^2_{\s(32)} + {\cal K}^1_{\s(32)}  \over s_{14} s_{3_\s 5} }
+ {{\cal K}^1_{\s(32)}  \over s_{2_\s 4} s_{51} } \cr
& \! \! \! \! \! \! \! \! \! \! \! \! \! \! \! \! \! \! \! \! \! \! \! \! \! \! \! +{-{\cal K}^3_{\s(32)} + {\cal K}^3_{\s(23)} - {\cal K}^1_{\s(23)} + {\cal K}^1_{\s(32)}  \over s_{2_\s 3_\s} s_{14} }
+ {-{\cal K}^2_{\s(32)} + {\cal K}^1_{\s(32)}  \over s_{3_\s 5} s_{2_\s 4} }
+ {-{\cal K}^1_{\s(23)} + {\cal K}^1_{\s(32)}  \over s_{51} s_{2_\s 3_\s} }
}}
The last pair of color orderings ${\cal A}_{5}(1,4,2_\s,3_\s,5)$ has more complicated numerators because of the coefficient 
$P_{(4,2_\s,3_\s)}{}^{(1,3_\s,2_\s)}= {1 \over s_{14} s_{3_\s 5}} + {\rm cyclic}(1,4,2_\s,3_\s,5)$ 
that addresses five different pole channels.

By comparing \fiveptmintwo\ with the ${\cal A}_5$ representation in the notation of \VamanP,
$$
{\cal A}_5(1,2_\rho,3_\rho,4_\rho,5) = { n [12_\rho,3_\rho,4_\rho 5] \over s_{12_\rho} s_{4_\rho 5}} + {\rm cyclic}(1,2_\rho,3_\rho,4_\rho,5),
$$ 
we can quickly read off the kinematic numerators:
\eqnn\vamaneco
\settabs \+ \hskip 8.4 cm & \hskip 4 cm &\cr
\+ $n[12_\s,3_\s,45] = {\cal K}^3_{\s(23)}$ & $n[3_\s 4,5,2_\s 1] = {\cal K}^2_{\s(23)}$ \cr
\+  $n[2_\s 3_\s,4,51] = {\cal K}^1_{\s(23)} -{\cal K}^1_{\s(32)}$ &  $n[2_\s 3_\s,1,45] = {\cal K}^3_{\s(32)} -{\cal K}^3_{\s(23)}$ \cr
\+ $n[3_\s 4,1,2_\s 5] = {\cal K}^1_{\s(23)} + {\cal K}^2_{\s(23)}$  & $n[12_\s,4,3_\s 5] = {\cal K}^3_{\s(23)} + {\cal K}^2_{\s(23)}$ \cr
\+ $n[14,3_\s,2_\s 5] = {\cal K}^3_{\s(32)} + {\cal K}^2_{\s(23)} + {\cal K}^2_{\s(32)} + {\cal K}^1_{\s(23)}$ & $n[51,2_\s,3_\s4] = {\cal K}^1_{\s(23)}$\cr
\+ $n[2_\s 3_\s,5,14] = {\cal K}^3_{\s(23)} -{\cal K}^3_{\s(32)} + {\cal K}^1_{\s(32)} -{\cal K}^1_{\s(23)}$ & \hfill \vamaneco & \cr
It is sufficient to display nine of them, the rest follows from $S_2$ relabelling $2 \leftrightarrow 3$. The $n_1,n_2 ,\ldots,n_{15}$ 
from the parametrization \fiveptKK\ translate into
\settabs 3 \columns
\+ $n_{1} = n[12,3,45]$ &  $n_{6} = n[14,3,25]$ & $ n_{11} = n[24,3,51]$ \cr
\+ $n_{2} = n[23,4,51]$ &  $n_{7} = n[32,5,14]$ & $ n_{12} = n[12,4,35]$ \cr
\+ $n_{3} = n[34,5,12]$ &  $n_{8} = n[25,1,43]$ & $ n_{13} = n[35,1,24]$ \cr
\+ $n_{4} = n[45,1,23]$ &  $n_{9} = n[13,4,25]$ & $ n_{14} = n[14,2,35]$ \cr
\+ $n_{5} = n[51,2,34]$ &  $n_{10} = n[42,5,13]$ & $ n_{15} = n[13,2,45]$ \cr
\noindent
The way in which the $n[ij,k,lm]$ are built out of ${\cal K}^j_{\s(2,3)}$ trivializes the Jacobi identities \Jacobifive\ or $n[ij,\{ k,lm \}] = 0$.
However, the expressions \vamaneco\ for $n[ij,k,lm]$ do not exhibit crossing symmetry including labels $1,4$ and $5$. 

In many instances, the symmetry properties \TrankNu\ of the BRST building blocks within ${\cal K}^l_\s$ allow to rewrite 
sums over several basic kinematics occurring in some $n_i$ as a single superfield, e.g.
$$
n_2= {\cal K}^1_{(23)} - {\cal K}^1_{(32)} = \langle (T_{123} - T_{132}) V_4 V_5 \rangle = \langle T_{321} V_4 V_5 \rangle.
$$
However, the right hand side is outside the five point basis of kinematics, so the Jacobi relations between numerators
are rather obscured by this building block manipulations. At any number of legs, the basis of ${\cal K}^l_\s$ is
designed such that all the symmetries of the building blocks are already exploited, so we refrain from performing 
manipulations like $T_{123} - T_{132}= T_{321}$ in higher order examples.

\subsec{Six-point}

In six-point amplitudes, the propagator matrix \defP\ can be completely constructed from the field theory limit of 
the four superstring subamplitudes associated with color orderings 
$\{1,2,3,4,5,6\}, \, \{1,2,3,5,4,6\}, \, \{1,2,5,3,4,6\}, \, \{1,5,2,3,4,6\}$. The $S_3$ relabelling covariance 
in $2,3,4$ connects them to the remaining 20 elements of the KK basis. Let us give some representative sample 
entries of $P_\rho{}^{(l,\s)}$ here,
\eqn \polesixpt{ \eqalign{
P_{(2345)}{}^{1,(423)} &= {1 \over s_{61} s_{23} s_{234}} , \ \ \ P_{(2345)}{}^{4,(432)} = {1 \over s_{56}s_{234}} \left( {1 \over s_{23}} + {1 \over s_{34}} \right) \cr
P_{(2345)}{}^{4,(234)} &=  {1 \over s_{56} }  \left( {1 \over s_{12} s_{34}} + {1 \over s_{12} s_{123}} + {1 \over s_{23} s_{123}} + {1 \over s_{23} s_{234}} + {1 \over s_{34} s_{234}} \right) \cr
P_{(2354)}{}^{3,(234)} &=  {1 \over s_{123} } \left( {1 \over s_{12}} + {1 \over s_{23}} \right) \left( {1 \over s_{45}} + {1 \over s_{46}} \right) \cr
P_{(5234)}{}^{1,(432)} &=  \frac{1}{s_{61} s_{25} s_{3  4 } } + \frac{1}{s_{15} s_{2  3 } s_{4 6} } + \frac{1}{s_{15} s_{12 5} s_{3  4 } } + \frac{1}{s_{12 5} s_{2 5} s_{34} } + \frac{1}{s_{15} s_{12 5} s_{ 46} } \cr
& + \ \frac{1}{s_{46} s_{25} s_{125} } + \frac{1}{s_{15} s_{23} s_{234} } + \frac{1}{s_{61} s_{23} s_{234} } + \frac{1}{s_{15} s_{234} s_{34} } + \frac{1}{s_{61} s_{234} s_{34} } \cr
&  + \ \frac{1}{s_{61} s_{23} s_{235} } + \frac{1}{s_{61} s_{25} s_{235} } + \frac{1}{s_{46} s_{23} s_{235} } + \frac{1}{s_{46} s_{25} s_{23 5} }
}}
and refer the reader to Appendix A for the complete result. 

Comparing the KK subamplitudes with
$$
{\cal A}_6(1,2_\rho,3_\rho,4_\rho,5_\rho,6) = { n [12_\rho,3_\rho 4_\rho, 5_\rho 6] \over s_{12_\rho} s_{3_\rho 4_\rho} s_{5_\rho 6}} 
+ { n [2_\rho3_\rho, 4_\rho 5_\rho ,61] \over s_{2_\rho 3_\rho} s_{4_\rho 5_\rho} s_{ 61}}
$$ 
$$
+ \left( {n[12_\rho,3_\rho,4_\rho,5_\rho 6] \over s_{12_\rho} s_{12_\rho 3_\rho} s_{5_\rho 6}} 
- { n[12_\rho,3_\rho,6,4_\rho 5_\rho] \over 2 s_{12_\rho} s_{12_\rho 3_\rho} s_{4_\rho 5_\rho}} 
- { n[2_\rho 3_\rho,1,4_\rho,5_\rho 6] \over 2 s_{2_\rho 3_\rho} s_{12_\rho 3_\rho} s_{5_\rho 6}} 
+ {\rm cyclic}(1,2_\rho,3_\rho,4_\rho,5_\rho,6) \right)
$$
allows to read off the 105 BCJ numerators. It is sufficient to display 
25 of them in $S_3$-covariant form:
\eqnn\sixpoint
\settabs \+ \hskip 3.1cm & \hskip 4.1 cm & \hskip 3.1cm & \hskip 5cm & \cr
\+ \hfill  $ n[12_\s,3_\s,4_\s,56]$ & $= {\cal K}^4_{\s(234)} $  & \hfill $ n[61,2_\s,3_\s, 5 4_\s]$ & $ =  {\cal K}^1_{\s(234)} $\cr
\+ \hfill  $ n[12_\s,3_\s,6,4_\s5]$ & $=  {\cal K}^3_{\s(234)} $ & \hfill $ n[12_\s,6,3_\s, 5 4_\s]$ & $=   {\cal K}^2_{\s(234)} $\cr
\+ \hfill  $ n[12_\s,3_\s4_\s,56]$ & $= {\cal K}^4_{\s(234)} - {\cal K}^4_{\s(243)} $ & \hfill $ n[2_\s3_\s,4_\s5,61]$ &$=   {\cal K}^1_{\s(324)} - {\cal K}^1_{\s(234)}$ \cr
\+ \hfill  $ n[12_\s,3_\s5, 6 4_\s]$ & $= {\cal K}^3_{\s(243)} + {\cal K}^2_{\s(243)} $ & \hfill $ n[3_\s4_\s,5,6,12_\s]$ &$=  {\cal K}^2_{\s(234)} - {\cal K}^2_{\s(243)} $\cr
\+ \hfill  $ n[4_\s5,6,1,2_\s3_\s]$ & $=  {\cal K}^3_{\s(324)}  - {\cal K}^3_{\s(234)}  $ & \hfill $ n[3_\s4_\s,5,2_\s,61]$ &$=  {\cal K}^1_{\s(234)} - {\cal K}^1_{\s(243)}$\cr
\+ \hfill  $ n[2_\s3_\s,1,4_\s,56]$ & $=  {\cal K}^4_{\s(324)} - {\cal K}^4_{\s(234)}  $  & \hfill $ n[12_\s,3_\s,5,4_\s6]$ &$=  {\cal K}^4_{\s(234)} + {\cal K}^3_{\s(234)} $\cr
\+ \hfill  $ n[4_\s6,1,2_\s,3_\s5]$ & $=  {\cal K}^2_{\s(423)} + {\cal K}^1_{\s(423)} $ & \cr
$$\eqalign{
    n[2_\s3_\s,4_\s,5,61] &=   - {\cal K}^1_{\s(234)} + {\cal K}^1_{\s(324)} + {\cal K}^1_{\s(423)} - {\cal K}^1_{\s(432)}\cr
    n[56,1,2_\s,3_\s4_\s] &=  {\cal K}^4_{\s(234)} - {\cal K}^4_{\s(243)} - {\cal K}^4_{\s(342)} + {\cal K}^4_{\s(432)}      \cr
    n[2_\s3_\s,5,1,4_\s6] &=  - {\cal K}^2_{\s(423)} + {\cal K}^2_{\s(432)} - {\cal K}^1_{\s(423)} + {\cal K}^1_{\s(432)} \cr
    n[2_\s3_\s,1,5,4_\s6] &=  {\cal K}^4_{\s(234)} - {\cal K}^4_{\s(324)} + {\cal K}^3_{\s(234)} - {\cal K}^3_{\s(324)} \cr
    n[12_\s,5,3_\s,4_\s6] &= {\cal K}^4_{\s(234)} + {\cal K}^3_{\s(234)} + {\cal K}^3_{\s(243)} + {\cal K}^2_{\s(243)} \cr
    n[3_\s4_\s,6,1,2_\s5] &= - {\cal K}^3_{\s(342)} + {\cal K}^3_{\s(432)} + {\cal K}^1_{\s(342)} - {\cal K}^1_{\s(432)} \cr
    n[12_\s,5,6,3_\s4_\s] &= - {\cal K}^4_{\s(234)} + {\cal K}^4_{\s(243)} + {\cal K}^2_{\s(234)} - {\cal K}^2_{\s(243)} \cr
    n[2_\s5,1,3_\s,4_\s6] &=  {\cal K}^3_{\s(342)} + {\cal K}^2_{\s(342)} + {\cal K}^2_{\s(432)} + {\cal K}^1_{\s(432)} 
}
$$
$$
\eqalign{
   n[15,2_\s,3_\s,4_\s6] &= {\cal K}^4_{\s(234)} + {\cal K}^3_{\s(234)} + {\cal K}^3_{\s(243)} + {\cal K}^3_{\s(342)} + {\cal K}^2_{\s(243)} + {\cal K}^2_{\s(342)} + {\cal K}^2_{\s(432)} + {\cal K}^1_{\s(432)}\cr
   n[2_\s3_\s,4_\s,6,15] &= {\cal K}^4_{\s(234)} - {\cal K}^4_{\s(324)} - {\cal K}^4_{\s(423)} + {\cal K}^4_{\s(432)} + {\cal K}^1_{\s(234)} - {\cal K}^1_{\s(324)} - {\cal K}^1_{\s(423)} + {\cal K}^1_{\s(432)} \cr
   n[15,2_\s,6,3_\s4_\s] &= {\cal K}^4_{\s(243)} - {\cal K}^4_{\s(234)} - {\cal K}^3_{\s(342)} + {\cal K}^3_{\s(432)} + {\cal K}^2_{\s(234)} - {\cal K}^2_{\s(243)} + {\cal K}^1_{\s(342)} - {\cal K}^1_{\s(432)} \cr
   n[15,2_\s3_\s,4_\s6] &= {\cal K}^4_{\s(234)} - {\cal K}^4_{\s(324)} + {\cal K}^3_{\s(234)} - {\cal K}^3_{\s(324)} - {\cal K}^2_{\s(423)} + {\cal K}^2_{\s(432)} - {\cal K}^1_{\s(423)} + {\cal K}^1_{\s(432)}
   }
$$
They have been explicitly checked to satisfy all the 105 Jacobi relations $n[ij,k, \{ l,mp \}]=0$ and 
$n[ij,kl,mp] = n[ij,k,l,mp] -n[ij,l,k,mp]$ (81 of which are linearly independent). It is interesting to 
note that the number of $K^l_\s$ forming the individual BCJ numerators is always a power of two, i.e. $1,2,4$ 
or $8$ in this case.

\subsec{Seven-point}

Since the number of channels grows like $(2n-5)!!$ in an $n$-point amplitude, a complete list of all BCJ numerators becomes 
very lengthy beyond six points. Appendix B gives all the 69 seven-point numerators which 
are not related by $2,3,4,5$ relabelling, they allow to obtain all the 945 numerators by going through 
the $\s \in S_4$ permutations of $(2,3,4,5)$. We also checked that all the 825 independent numerators 
equations (out of 1260 in total) are satisfied.

\subsec{Higher-point and general observations}

Instead of continuing the numerator list to higher points, we conclude this section with some general 
remarks and observation on the structure of the string inspired expressions for the BCJ numerators.

Firstly, entries of the $n$ point propagator matrix always factorize into sums of $m$ propagators with $C(m) = {(2m)! \over m! (m+1)!}$ terms,
\eqn \catalan{
P_\rho{}^{(l,\s)} \sim  \prod \left( \sum_{j=1}^{C(m)} {1 \over s_{(\alpha^1)_{j}}
s_{(\alpha^2)_j} \ldots s_{(\alpha^{m})_j}} \right)
}
where $C(m)$ is the $m$'th Catalan number and counts the number of channels appearing in a $m+2$ point color 
ordered amplitude \NimaCatalan. At $n=5$, we have seen three different pole structures in \fiveptmintwo,
$$
P_\rho{}^{(l,\s)} \Big. \Big|_{n=5} \sim \left\{ {1 \over s_\alpha s_\beta}, \, {1 \over s_\alpha } 
\left( { 1 \over s_{\beta_1}} + {1 \over s_{\beta_2}} \right) , \,  \sum_{i=1}^5{ 1 \over s_{\alpha_i}s_{ \beta_i} } \right\}
$$
and the six point analogue contains the five types of products displayed in \polesixpt:
$$
P_\rho{}^{(l,\s)} \Big. \Big|_{n=6} \sim \left\{ {1 \over s_\alpha s_\beta s_\gamma}, \, {1 \over s_\alpha s_\beta} 
\left( { 1 \over s_{\gamma_1}} + {1 \over s_{ \gamma_2}} \right), \, {1 \over s_\alpha } \left( { 1 \over s_{\beta_1}} 
+ {1 \over s_{\beta_2}} \right) \left( { 1 \over s_{\gamma_1}} + {1 \over s_{\gamma_2}} \right) ,\right.
$$
$$
\left. {1 \over s_\alpha} \sum_{i=1}^5{ 1 \over s_{\beta_i}s_{ \gamma_i} } , \, \sum_{j=1}^{14}{ 1 \over s_{\alpha_j} s_{\beta_j}s_{ \gamma_j} } \right\}
$$
The pattern was observed to persist up to eight-point. However, not all possible partitions of the overall $n-3$ 
propagators into products of type \catalan\ are realized. For instance, there are no terms like 
$\left( {1 \over s_{\alpha_1}} + {1 \over s_{\alpha_2}} \right) \left( {1 \over s_{\beta_1}} + {1 \over s_{\beta_2}} \right)$ 
at five points, $\left( {1 \over s_{\alpha_1}} + {1 \over s_{\alpha_2}} \right) \left( \sum_{i=1}^5 {1 \over s_{\beta_i} s_{\gamma_i}} \right)$ 
at six points and $\left( {1 \over s_{\alpha_1}} + {1 \over s_{\alpha_2}} \right) \left( {1 \over s_{\beta_1}} 
+ {1 \over s_{\beta_2}} \right)\left( {1 \over s_{\gamma_1}} + {1 \over s_{\gamma_2}} \right) { 1 \over s_\delta}$ at seven points.

Secondly, the number of ${\cal K}^l_\s$ kinematics entering the individual $n$ point BCJ numerators up to $n=8$ 
is always a power of two $1,2,4,\ldots , 2^{n-3}$. This can be largely explained from the flipping antisymmetry 
of $n[ij,k,\ldots]$ in pairs of labels $i,j$ sharing a terminal three point vertex. If they are both from the 
range $i_\s,j_\s \in \{ 2,3,\ldots,n-2 \}$, then the  ${\cal K}^l_\s$ are required to pair up with their 
$i \leftrightarrow j$ images. Moreover, if several other $2,3,\ldots,n-2$ labels $k_\s,l_\s$ follow, then 
a nested antisymmetrization emerges, e.g. $n[i_\s j_\s,k_\s,l_\s,\ldots] \leftrightarrow {\cal K}^l_{\s( [[[i_\s j_\s] k_\s ]l_\s] \ldots)}$.

Another source of doubling the terms is a terminal vertex with legs $1$ and $n-1$: Swapping 
$1\leftrightarrow n-1$ maps ${\cal K}^l_\s$ to ${\cal K}^{n-1-l}_{\bar \sigma}$ where $\bar \sigma$ 
denotes the permutation of reverse order, $\bar \sigma(23\ldots p-1,p) = \sigma(p,p-1,\ldots32)$. That 
is why $n[1(n-1),\ldots]$ can only contain pairs like ${\cal K}^l_\s + {\cal K}^{n-1-l}_{\bar \sigma}$ 
which might be further antisymmetrized in some legs from $\{ 2,3,\ldots,n-2 \}$ due to another terminal vertex.

The following table shows the distribution of the $(2n-5)!!$ numerators into 
packages of $2^j$ basis elements:
\bigskip
\moveright 2 cm
\vbox{\offinterlineskip
\halign{\strut \vrule \hfil\quad # \quad\hfil &\vrule\hskip 0.02cm \vrule \hfil \quad # \quad \hfil 
& \vrule \hfil \quad # \quad \hfil 
& \vrule \hfil \quad # \quad \hfil 
& \vrule \hfil \quad # \quad \hfil 
& \vrule \hfil \quad # \quad \hfil \vrule \cr
\noalign{\hrule}
\# terms & 4pts & 5pts & 6pts & 7pts & 8pts \cr
\noalign{\hrule\vskip 0.02cm \hrule}
1 & 2 & 6 & 24 & 120 & 720 \cr 
\noalign{\hrule}
2 & 1 & 6 & 36 & 240 & 1800 \cr 
\noalign{\hrule}
4 &  & 3 & 30 & 270 & 2520 \cr 
\noalign{\hrule}
8 &  &  & 15 & 210 & 2520 \cr 
\noalign{\hrule}
16 &  &  &  & 105 & 1890 \cr 
\noalign{\hrule}
32 &  &  &  &  & 945 \cr 
\noalign{\hrule}
}}
{\leftskip=40pt\rightskip=40pt\noindent\font\smallrm=cmr6
{\bf Table 1.} The number of BCJ numerators in $n$-point amplitudes containing $2^j$ basis kinematics ${\cal K}^l_\s$ for $j=0,1,{\ldots},n-3$.
\par }
\bigskip
\noindent
We suspect that the grading of kinematic numerators according to their ${\cal K}^l_\s$ content is 
connected with the factorization pattern \catalan\ of $P_\rho{}^{(l,\s)}$ entries.

\newsec{Concluding remarks}

In this paper we have developed a method based on string theory to construct kinematic factors $n_i$ for 
gauge theory amplitudes which manifestly obey Jacobi identities dual to the color algebra $c_i + c_j + c_k=0$. The fact 
that the vanishing of the dual numerator triplet $n_i+n_j+n_k$ depends on the organization of contact terms 
complicates the direct construction of $n_i$ within the gauge theory setup.

The pure spinor approach to superstring theory naturally introduces a kinematic basis \Kbasic\ of $(n-2)!$ 
elements for $n$-point tree amplitudes of the gauge multiplet. The field theory amplitude can be extracted by taking 
the low energy limit of the string result \stringBCJ\ using the method of \WIP. This determines the BCJ 
numerators $n_i$ for any pole channel in term of the $(n-2)!$ basis kinematics. The resulting expressions for the 
$n_i$ are manifestly local and supersymmetric. Although they originate from the ten dimensional SYM theory, 
it is still straightforward to dimensionally reduce the superfield components and to recycle the purely 
bosonic amplitudes for QCD or any other theory with less than sixteen supercharges. 

The basis dimension $(n-2)!$ together with the BCJ relations between color ordered field theory amplitudes 
imply that the string inspired $n_i$ satisfy the dual Jacobi identities $n_i+n_j+n_k=0$ for each vanishing triplet 
of color factors $(c_i,c_j,c_k)$. However, as a price to pay for the exact $n_i \leftrightarrow c_i$ duality, crossing symmetry 
is lost for the kinematic numerators. This can be immediately recognized from the explicit solutions \vamaneco, 
\sixpoint\ and appendix B for the $n_i$ at five, six and seven points, respectively. It would be interesting 
to find a compact form for crossing symmetric numerators while still preserving the Jacobi-like relations.

\goodbreak
\vskip 5mm
\centerline{\noindent{\bf Acknowledgments} }

We thank the Kavli Institute for Theoretical Physics in Santa Barbara 
for hospitality and financial support during preparation and completion of this work.
We are indebted to Zvi Bern for initiating this work and for helpful discussions. Furthermore, we are grateful to John Joseph Carrasco and Henrik Johansson for strong encouragement, stimulating discussions and for contributing valuable suggestions to the manuscript. This research was supported in part by the National Science Foundation under Grant No. NSF PHY05--51164. 
C.M. thanks the partial financial support from
the MPG and acknowledges support by the Deutsch-Israelische
Projektkooperation (DIP H52).



\appendix{A}{Field theory limit of six-point integrals}

This appendix contains the field theory limit of the six-point superstring amplitudes in the KK color orderings. It is the higher point analogue of the five point result \fiveptmintwo.
$$
   {\cal A}(1,2_\s,3_\s,4_\s,5,6) =
   { {\cal K}^3_{\s(234)} \over    s_{12_\s} s_{4_\s 5} s_{12_\s 3_\s}  }
  + {  - {\cal K}^2_{\s(234)} \over    s_{12_\s} s_{4_\s 5} s_{3_\s 4_\s 5}  } 
$$
$$
  + {  - {\cal K}^4_{\s(234)} \over    s_{12_\s} s_{56} s_{12_\s 3_\s}  }
  + { {\cal K}^1_{\s(234)} \over    s_{16} s_{4_\s 5} s_{3_\s 4_\s 5}  }
  + {  - {\cal K}^4_{\s(234)} + {\cal K}^4_{\s(243)} \over    s_{12_\s} s_{3_\s 4_\s} s_{56}  }
  + {  - {\cal K}^2_{\s(234)} + {\cal K}^2_{\s(243)} \over    s_{12_\s} s_{3_\s 4_\s} s_{3_\s 4_\s 5}  }
$$
$$
  + { {\cal K}^1_{\s(234)} - {\cal K}^1_{\s(324)} \over    s_{16} s_{2_\s 3_\s} s_{4_\s 5}  }
  + { {\cal K}^1_{\s(234)} - {\cal K}^1_{\s(243)} \over    s_{16} s_{3_\s 4_\s} s_{3_\s 4_\s 5}  }
  + { {\cal K}^3_{\s(234)} - {\cal K}^3_{\s(324)} \over    s_{2_\s 3_\s} s_{4_\s 5} s_{12_\s 3_\s}  }
  + {  - {\cal K}^4_{\s(234)} + {\cal K}^4_{\s(324)} \over    s_{2_\s 3_\s} s_{56} s_{12_\s 3_\s}  }
$$
$$
  + { {\cal K}^1_{\s(234)} - {\cal K}^1_{\s(324)} - {\cal K}^1_{\s(423)} + {\cal K}^1_{\s(432)} \over    s_{16} s_{2_\s 3_\s} s_{2_\s 3_\s 4_\s}  }
  + { {\cal K}^1_{\s(234)} - {\cal K}^1_{\s(243)} - {\cal K}^1_{\s(342)} + {\cal K}^1_{\s(432)} \over    s_{16} s_{3_\s 4_\s} s_{2_\s 3_\s 4_\s}  }
$$
$$
  + {  - {\cal K}^4_{\s(234)} + {\cal K}^4_{\s(324)} + {\cal K}^4_{\s(423)} - {\cal K}^4_{\s(432)} \over    s_{2_\s 3_\s} s_{56} s_{2_\s 3_\s 4_\s}  }
  + {  - {\cal K}^4_{\s(234)} + {\cal K}^4_{\s(243)} + {\cal K}^4_{\s(342)} - {\cal K}^4_{\s(432)} \over    s_{3_\s 4_\s} s_{56} s_{2_\s 3_\s 4_\s}  }
$$

$$   {\cal A}(1,2_\s,3_\s,5,4_\s,6) =
   { {\cal K}^2_{\s(243)} \over    s_{12_\s} s_{3_\s 5} s_{3_\s 4_\s 5}  } 
  + {  - {\cal K}^3_{\s(234)} \over    s_{12_\s} s_{4_\s 5} s_{12_\s 3_\s}  }
  + { {\cal K}^2_{\s(234)} \over    s_{12_\s} s_{4_\s 5} s_{3_\s 4_\s 5}  }$$
$$  + {  - {\cal K}^1_{\s(423)} \over    s_{16} s_{3_\s 5} s_{2_\s 3_\s 5}  }
  + {  - {\cal K}^1_{\s(243)} \over    s_{16} s_{3_\s 5} s_{3_\s 4_\s 5}  }
  + {  - {\cal K}^1_{\s(234)} \over    s_{16} s_{4_\s 5} s_{3_\s 4_\s 5}  }
  + { {\cal K}^3_{\s(243)} + {\cal K}^2_{\s(243)} \over    s_{12_\s} s_{3_\s 5} s_{4_\s 6}  }
  + {  - {\cal K}^4_{\s(234)} - {\cal K}^3_{\s(234)} \over    s_{12_\s} s_{4_\s 6} s_{12_\s 3_\s}  }$$
$$  + {  - {\cal K}^1_{\s(234)} + {\cal K}^1_{\s(324)} \over    s_{16} s_{2_\s 3_\s} s_{4_\s 5}  }
  + {  - {\cal K}^1_{\s(423)} + {\cal K}^1_{\s(432)} \over    s_{16} s_{2_\s 3_\s} s_{2_\s 3_\s 5}  }
  + {  - {\cal K}^3_{\s(234)} + {\cal K}^3_{\s(324)} \over    s_{2_\s 3_\s} s_{4_\s 5} s_{12_\s 3_\s}  }
  + {  - {\cal K}^2_{\s(423)} - {\cal K}^1_{\s(423)} \over    s_{3_\s 5} s_{4_\s 6} s_{2_\s 3_\s 5}  }$$
$$  + {  - {\cal K}^4_{\s(234)} + {\cal K}^4_{\s(324)} - {\cal K}^3_{\s(234)} + {\cal K}^3_{\s(324)} \over    s_{2_\s 3_\s} s_{4_\s 6} s_{12_\s 3_\s}  }
  + {  - {\cal K}^2_{\s(423)} + {\cal K}^2_{\s(432)} - {\cal K}^1_{\s(423)} + {\cal K}^1_{\s(432)} \over    s_{2_\s 3_\s} s_{4_\s 6} s_{2_\s 3_\s 5}  }$$

$$   {\cal A}(1,2_\s,5,3_\s,4_\s,6) =
   {  - {\cal K}^2_{\s(243)} \over    s_{12_\s} s_{3_\s 5} s_{3_\s 4_\s 5}  }
  + { {\cal K}^1_{\s(432)} \over    s_{16} s_{2_\s 5} s_{2_\s 3_\s 5}  }$$
$$  + { {\cal K}^1_{\s(423)} \over    s_{16} s_{3_\s 5} s_{2_\s 3_\s 5}  }
  + { {\cal K}^1_{\s(243)} \over    s_{16} s_{3_\s 5} s_{3_\s 4_\s 5}  }
  + {  - {\cal K}^1_{\s(342)} + {\cal K}^1_{\s(432)} \over    s_{16} s_{2_\s 5} s_{3_\s 4_\s}  }
  + {  - {\cal K}^1_{\s(234)} + {\cal K}^1_{\s(243)} \over    s_{16} s_{3_\s 4_\s} s_{3_\s 4_\s 5}  }$$
$$  + { {\cal K}^2_{\s(234)} - {\cal K}^2_{\s(243)} \over    s_{12_\s} s_{3_\s 4_\s} s_{3_\s 4_\s 5}  }
  + {  - {\cal K}^3_{\s(243)} - {\cal K}^2_{\s(243)} \over    s_{12_\s} s_{3_\s 5} s_{4_\s 6}  }
  + { {\cal K}^2_{\s(432)} + {\cal K}^1_{\s(432)} \over    s_{2_\s 5} s_{4_\s 6} s_{2_\s 3_\s 5}  }
  + { {\cal K}^2_{\s(423)} + {\cal K}^1_{\s(423)} \over    s_{3_\s 5} s_{4_\s 6} s_{2_\s 3_\s 5}  }$$
$$  + {  - {\cal K}^4_{\s(234)} - {\cal K}^3_{\s(234)} - {\cal K}^3_{\s(243)} - {\cal K}^2_{\s(243)} \over    s_{12_\s} s_{4_\s 6} s_{12_\s 5}  }
  + {  - {\cal K}^4_{\s(234)} + {\cal K}^4_{\s(243)} + {\cal K}^2_{\s(234)} - {\cal K}^2_{\s(243)} \over    s_{12_\s} s_{3_\s 4_\s} s_{12_\s 5}  }$$
$$  + { {\cal K}^3_{\s(342)} - {\cal K}^3_{\s(432)} - {\cal K}^1_{\s(342)} + {\cal K}^1_{\s(432)} \over    s_{2_\s 5} s_{3_\s 4_\s} s_{12_\s 5}  }
  + { {\cal K}^3_{\s(342)} + {\cal K}^2_{\s(342)} + {\cal K}^2_{\s(432)} + {\cal K}^1_{\s(432)} \over    s_{2_\s 5} s_{4_\s 6} s_{12_\s 5}  }$$

$$   {\cal A}(1,5,2_\s,3_\s,4_\s,6) =
   {  - {\cal K}^3_{\s(342)} - {\cal K}^2_{\s(342)} - {\cal K}^2_{\s(432)} - {\cal K}^1_{\s(432)} \over    s_{2_\s 5} s_{4_\s 6} s_{12_\s 5}  }$$
$$  + {  - {\cal K}^4_{\s(234)} + {\cal K}^4_{\s(324)} - {\cal K}^3_{\s(234)} + {\cal K}^3_{\s(324)} + {\cal K}^2_{\s(423)} - {\cal K}^2_{\s(432)} + {\cal K}^1_{\s(423)} - {\cal K}^1_{\s(432)} \over    s_{15} s_{2_\s 3_\s} s_{4_\s 6}  }$$
$$  + {  - {\cal K}^4_{\s(234)} + {\cal K}^4_{\s(324)} + {\cal K}^4_{\s(423)} - {\cal K}^4_{\s(432)} - {\cal K}^1_{\s(234)} + {\cal K}^1_{\s(324)} + {\cal K}^1_{\s(423)} - {\cal K}^1_{\s(432)} \over    s_{15} s_{2_\s 3_\s} s_{2_\s 3_\s 4_\s}  }$$
$$  + {  - {\cal K}^4_{\s(234)} + {\cal K}^4_{\s(243)} - {\cal K}^3_{\s(342)} + {\cal K}^3_{\s(432)} + {\cal K}^2_{\s(234)} - {\cal K}^2_{\s(243)} + {\cal K}^1_{\s(342)} - {\cal K}^1_{\s(432)} \over    s_{15} s_{3_\s 4_\s} s_{12_\s 5}  }$$
$$  + {  - {\cal K}^4_{\s(234)} + {\cal K}^4_{\s(243)} + {\cal K}^4_{\s(342)} - {\cal K}^4_{\s(432)} - {\cal K}^1_{\s(234)} + {\cal K}^1_{\s(243)} + {\cal K}^1_{\s(342)} - {\cal K}^1_{\s(432)} \over    s_{15} s_{3_\s 4_\s} s_{2_\s 3_\s 4_\s}  }$$
$$  + {  - {\cal K}^4_{\s(234)} - {\cal K}^3_{\s(234)} - {\cal K}^3_{\s(243)} - {\cal K}^3_{\s(342)} - {\cal K}^2_{\s(243)} - {\cal K}^2_{\s(342)} - {\cal K}^2_{\s(432)} - {\cal K}^1_{\s(432)} \over    s_{15} s_{4_\s 6} s_{12_\s 5}  }$$
$$  + {  - {\cal K}^1_{\s(432)} \over    s_{16} s_{2_\s 5} s_{2_\s 3_\s 5}  }
  + { {\cal K}^1_{\s(423)} - {\cal K}^1_{\s(432)} \over    s_{16} s_{2_\s 3_\s} s_{2_\s 3_\s 5}  }
  + { {\cal K}^1_{\s(342)} - {\cal K}^1_{\s(432)} \over    s_{16} s_{2_\s 5} s_{3_\s 4_\s}  }
  + {  - {\cal K}^2_{\s(432)} - {\cal K}^1_{\s(432)} \over    s_{2_\s 5} s_{4_\s 6} s_{2_\s 3_\s 5}  }$$
$$  + {  - {\cal K}^1_{\s(234)} + {\cal K}^1_{\s(324)} + {\cal K}^1_{\s(423)} - {\cal K}^1_{\s(432)} \over    s_{16} s_{2_\s 3_\s} s_{2_\s 3_\s 4_\s}  }
  + {  - {\cal K}^1_{\s(234)} + {\cal K}^1_{\s(243)} + {\cal K}^1_{\s(342)} - {\cal K}^1_{\s(432)} \over    s_{16} s_{3_\s 4_\s} s_{2_\s 3_\s 4_\s}  }$$
$$  + { {\cal K}^2_{\s(423)} - {\cal K}^2_{\s(432)} + {\cal K}^1_{\s(423)} - {\cal K}^1_{\s(432)} \over    s_{2_\s 3_\s} s_{4_\s 6} s_{2_\s 3_\s 5}  }
  + {  - {\cal K}^3_{\s(342)} + {\cal K}^3_{\s(432)} + {\cal K}^1_{\s(342)} - {\cal K}^1_{\s(432)} \over    s_{2_\s 5} s_{3_\s 4_\s} s_{12_\s 5}  }$$
Note that the basis kinematics ${\cal K}^1_{\s(432)}$ contributes to each of the fourteen pole channels of ${\cal A}_6(1,5,2_\s,3_\s,4_\s,6)$.

\appendix{B}{Seven-point numerators}

This appendix lists the kinematic numerators for the seven-point amplitude. Thanks to $S_4$ covariance in $(2,3,4,5)$ permutations, only 69 out of the 945 BCJ numerators have to be displayed explicitly, the rest follows from relabelling of legs $2,3,4$ and $5$:
\eqnn\sevenpa
\settabs \+ \hskip 3.1cm & \hskip 4.1 cm & \hskip 3.1cm \hskip 0.1cm & \cr
\+ \hfill  $ n[1 2_\s , 3_\s , 4_\s , 5_\s ,67]$ & $=  {\cal K}^5_{\s(2345)} $  & $ n[ 5_\s 6, 4_\s ,7, 3_\s ,1 2_\s ]$ & $=  {\cal K}^3_{\s(2345)} $ \hfil \cr
\+ \hfill  $ n[71, 2_\s , 3_\s , 4_\s , 5_\s 6]$ & $=  {\cal K}^1_{\s(2345)} $  & $ n[1 2_\s , 3_\s , 4_\s ,7, 5_\s 6]$ & $=  {\cal K}^4_{\s(2345)} $ \hfill \cr
\+ \hfill  $ n[1 2_\s ,7, 3_\s , 4_\s , 5_\s 6]$ & $=  {\cal K}^2_{\s(2345)} $ \cr
%
\settabs \+ \hskip 3.1cm & \hskip 4.1 cm & \hskip 3.1cm & \hskip 5cm & \cr
\+ \hfill  $ n[1 2_\s , 3_\s , 4_\s  5_\s ,67]$ & $= {\cal K}^5_{\s(2345)} - {\cal K}^5_{\s(2354)}$ & \hfill $ n[ 4_\s 6, 3_\s , 5_\s 7, 2_\s 1]$ & $=  {\cal K}^2_{\s(2534)} + {\cal K}^3_{\s(2534)}$\cr
\+ \hfill $ n[1 2_\s ,7, 3_\s ,6, 4_\s  5_\s ]$ & $=  {\cal K}^2_{\s(2354)} - {\cal K}^2_{\s(2345)}$ & \hfill $ n[ 2_\s  3_\s ,1, 4_\s ,7, 5_\s 6]$ & $=  {\cal K}^4_{\s(3245)} - {\cal K}^4_{\s(2345)} $ \cr
\+ \hfill $ n[ 5_\s 6,7,1 2_\s , 3_\s  4_\s ]$  & $=  {\cal K}^4_{\s(2435)} - {\cal K}^4_{\s(2345)} $ & \hfill $ n[71, 2_\s , 3_\s  4_\s , 5_\s 6]$ & $=  {\cal K}^1_{\s(2345)} - {\cal K}^1_{\s(2435)} $\cr
\+ \hfill $ n[1 2_\s , 3_\s , 4_\s 6, 7 5_\s ]$ & $=  {\cal K}^3_{\s(2354)} + {\cal K}^4_{\s(2354)}$ & \hfill $ n[1 2_\s , 3_\s , 4_\s ,6, 5_\s 7]$ & $=  {\cal K}^4_{\s(2345)} + {\cal K}^5_{\s(2345)} $\cr
\+ \hfill $ n[1 2_\s ,7, 3_\s  4_\s , 5_\s 6]$  & $=  {\cal K}^2_{\s(2345)} - {\cal K}^2_{\s(2435)}$ & \hfill $ n[ 2_\s  3_\s ,1, 4_\s , 5_\s ,67]$ & $=  {\cal K}^5_{\s(3245)} - {\cal K}^5_{\s(2345)}  $ \cr
\+ \hfill $ n[ 4_\s  5_\s ,6,7, 3_\s ,1 2_\s ]$ & $=  {\cal K}^3_{\s(2354)} - {\cal K}^3_{\s(2345)} $ & \hfill $ n[ 5_\s 6, 4_\s ,71, 2_\s  3_\s ]$  & $=  {\cal K}^1_{\s(3245)} - {\cal K}^1_{\s(2345)}  $\cr
\+ \hfill $ n[ 5_\s 6, 4_\s ,7,1, 2_\s  3_\s ]$ & $=  {\cal K}^3_{\s(3245)} - {\cal K}^3_{\s(2345)} $ & \hfill $ n[ 5_\s 7,1, 2_\s , 3_\s , 6 4_\s ]$ & $=  {\cal K}^1_{\s(5234)} + {\cal K}^2_{\s(5234)} $\cr
\+ \hfill $ n[67, 5_\s ,1 2_\s , 3_\s  4_\s ]$  & $=  {\cal K}^5_{\s(2435)} - {\cal K}^5_{\s(2345)} $ & \hfill $ n[71, 2_\s , 3_\s ,6, 4_\s  5_\s ]$ & $ =  {\cal K}^1_{\s(2354)} - {\cal K}^1_{\s(2345)}  $\cr

\settabs \+ \hskip 3.1cm & \cr
\+ \hfill $ n[ 4_\s  5_\s ,6,71, 2_\s  3_\s ]$ & $= {\cal K}^1_{\s(2345)} - {\cal K}^1_{\s(2354)} - {\cal K}^1_{\s(3245)} + {\cal K}^1_{\s(3254)} $\cr
\+ \hfill $ n[ 4_\s  5_\s ,6,7,1, 2_\s  3_\s ]$ & $=  {\cal K}^3_{\s(2345)} - {\cal K}^3_{\s(2354)} - {\cal K}^3_{\s(3245)} + {\cal K}^3_{\s(3254)} $\cr
\+ \hfill $ n[ 2_\s  3_\s ,1, 4_\s  5_\s ,67]$ & $=  {\cal K}^5_{\s(2354)} - {\cal K}^5_{\s(2345)}  + {\cal K}^5_{\s(3245)} - {\cal K}^5_{\s(3254)} $\cr
\+ \hfill $ n[ 2_\s  3_\s , 4_\s , 5_\s 6,71]$ & $=  {\cal K}^1_{\s(2345)} - {\cal K}^1_{\s(3245)} - {\cal K}^1_{\s(4235)} + {\cal K}^1_{\s(4325)} $\cr
\+ \hfill $ n[ 3_\s  4_\s , 5_\s ,6,7,1 2_\s ]$ & $=  {\cal K}^2_{\s(2435)} - {\cal K}^2_{\s(2345)} + {\cal K}^2_{\s(2534)} - {\cal K}^2_{\s(2543)} $\cr
\+ \hfill $ n[ 3_\s  4_\s , 5_\s ,67,1 2_\s ]$ & $=  {\cal K}^5_{\s(2345)} - {\cal K}^5_{\s(2435)} - {\cal K}^5_{\s(2534)} + {\cal K}^5_{\s(2543)} $\cr
\+ \hfill $ n[ 3_\s  4_\s ,6, 5_\s 7,1 2_\s ]$ & $=  {\cal K}^2_{\s(2534)} - {\cal K}^2_{\s(2543)} + {\cal K}^3_{\s(2534)} - {\cal K}^3_{\s(2543)} $\cr
\+ \hfill $ n[ 4_\s  5_\s ,7,1, 2_\s , 3_\s 6]$ & $=  {\cal K}^1_{\s(5423)} - {\cal K}^1_{\s(4523)} + {\cal K}^3_{\s(4523)} - {\cal K}^3_{\s(5423)} $\cr
\+ \hfill $ n[ 4_\s  5_\s ,7,1 2_\s , 3_\s 6]$ & $=  {\cal K}^2_{\s(2453)} - {\cal K}^2_{\s(2543)} - {\cal K}^4_{\s(2453)} + {\cal K}^4_{\s(2543)} $\cr
\+ \hfill $ n[ 5_\s 6,7,1, 2_\s , 3_\s  4_\s ]$ & $=  {\cal K}^4_{\s(2435)} - {\cal K}^4_{\s(2345)} + {\cal K}^4_{\s(3425)} - {\cal K}^4_{\s(4325)} $\cr
\+ \hfill $ n[ 5_\s 7,1, 2_\s  3_\s , 4_\s 6]$ & $=  {\cal K}^1_{\s(5324)} - {\cal K}^1_{\s(5234)} - {\cal K}^2_{\s(5234)} + {\cal K}^2_{\s(5324)} $\cr
\+ \hfill $ n[1 2_\s , 3_\s ,6, 4_\s , 5_\s 7]$ & $=  {\cal K}^3_{\s(2354)} + {\cal K}^4_{\s(2345)} + {\cal K}^4_{\s(2354)} + {\cal K}^5_{\s(2345)} $\cr
\+ \hfill $ n[1 2_\s , 3_\s ,6,7, 4_\s  5_\s ]$ & $=  {\cal K}^3_{\s(2345)} - {\cal K}^3_{\s(2354)} - {\cal K}^5_{\s(2345)} + {\cal K}^5_{\s(2354)} $\cr
\+ \hfill $ n[ 2_\s  3_\s ,1, 4_\s 6, 5_\s 7]$ & $=  {\cal K}^3_{\s(2354)} - {\cal K}^3_{\s(3254)} + {\cal K}^4_{\s(2354)} - {\cal K}^4_{\s(3254)} $\cr
\+ \hfill $ n[ 2_\s  3_\s ,1, 4_\s ,6, 5_\s 7]$ & $=  {\cal K}^4_{\s(3245)} - {\cal K}^4_{\s(2345)}  - {\cal K}^5_{\s(2345)} + {\cal K}^5_{\s(3245)} $\cr
\+ \hfill $ n[ 3_\s  4_\s , 5_\s ,6, 2_\s ,71]$ & $=  {\cal K}^1_{\s(2435)} - {\cal K}^1_{\s(2345)} + {\cal K}^1_{\s(2534)} - {\cal K}^1_{\s(2543)} $\cr
\+ \hfill $ n[ 5_\s 7,1, 2_\s ,6, 3_\s  4_\s ]$ & $=  {\cal K}^1_{\s(5234)} - {\cal K}^1_{\s(5243)} + {\cal K}^2_{\s(5234)} - {\cal K}^2_{\s(5243)} $\cr
\+ \hfill $ n[ 5_\s 7,6,1 2_\s , 3_\s  4_\s ]$ & $=  {\cal K}^4_{\s(2435)} - {\cal K}^4_{\s(2345)} - {\cal K}^5_{\s(2345)} + {\cal K}^5_{\s(2435)} $\cr
\+ \hfill $ n[67, 5_\s ,1, 2_\s , 3_\s  4_\s ]$ & $=  {\cal K}^5_{\s(2435)} - {\cal K}^5_{\s(2345)} + {\cal K}^5_{\s(3425)} - {\cal K}^5_{\s(4325)} $\cr
\+ \hfill $ n[ 5_\s 7, 4_\s ,1, 2_\s , 6 3_\s ]$ & $=  {\cal K}^1_{\s(5423)} + {\cal K}^2_{\s(4523)} + {\cal K}^2_{\s(5423)} + {\cal K}^3_{\s(4523)} $\cr
\+ \hfill $ n[ 5_\s 7, 4_\s ,1 2_\s , 3_\s 6]$ & $=  {\cal K}^2_{\s(2543)} + {\cal K}^3_{\s(2453)} + {\cal K}^3_{\s(2543)} + {\cal K}^4_{\s(2453)} $\cr

\settabs \+ \hskip 3.1cm & \cr
\+ \hfill $ n[ 2_\s  3_\s ,1,6,7, 4_\s  5_\s ]$ & $= - {\cal K}^3_{\s(2345)} + {\cal K}^3_{\s(2354)} + {\cal K}^3_{\s(3245)} - {\cal K}^3_{\s(3254)}  $ \cr
\+ & \quad $ + {\cal K}^5_{\s(2345)} - {\cal K}^5_{\s(2354)} - {\cal K}^5_{\s(3245)} + {\cal K}^5_{\s(3254)} $ \cr
\+ \hfill $ n[67,1, 2_\s  3_\s , 4_\s  5_\s ]$ & $= + {\cal K}^5_{\s(2345)} - {\cal K}^5_{\s(2354)} - {\cal K}^5_{\s(3245)} + {\cal K}^5_{\s(3254)} $ \cr
\+ & \quad $ - {\cal K}^5_{\s(4523)} + {\cal K}^5_{\s(4532)} + {\cal K}^5_{\s(5423)} - {\cal K}^5_{\s(5432)} $ \cr
\+ \hfill $ n[1 2_\s ,6, 3_\s  4_\s , 5_\s 7]$ & $= - {\cal K}^2_{\s(2534)} + {\cal K}^2_{\s(2543)} - {\cal K}^3_{\s(2534)} + {\cal K}^3_{\s(2543)} $ \cr
\+ & \quad $ + {\cal K}^4_{\s(2345)} - {\cal K}^4_{\s(2435)} + {\cal K}^5_{\s(2345)} - {\cal K}^5_{\s(2435)} $\cr
\+ \hfill $ n[ 2_\s  3_\s , 4_\s , 5_\s ,6,71]$ & $= + {\cal K}^1_{\s(2345)} - {\cal K}^1_{\s(3245)} - {\cal K}^1_{\s(4235)} + {\cal K}^1_{\s(4325)} $ \cr
\+ & \quad $ - {\cal K}^1_{\s(5234)} + {\cal K}^1_{\s(5324)} + {\cal K}^1_{\s(5423)} - {\cal K}^1_{\s(5432)} $\cr
\+ \hfill $ n[ 2_\s 6,1, 3_\s ,7, 4_\s  5_\s ]$ & $= + {\cal K}^1_{\s(4532)} - {\cal K}^1_{\s(5432)} + {\cal K}^2_{\s(3452)} - {\cal K}^2_{\s(3542)} $ \cr
\+ & \quad $ - {\cal K}^3_{\s(4532)} + {\cal K}^3_{\s(5432)} - {\cal K}^4_{\s(3452)} + {\cal K}^4_{\s(3542)} $\cr     
\+ \hfill $ n[ 3_\s  4_\s , 5_\s ,7,1, 2_\s 6]$ & $= - {\cal K}^1_{\s(3452)} + {\cal K}^1_{\s(4352)} + {\cal K}^1_{\s(5342)} - {\cal K}^1_{\s(5432)} $ \cr
\+ & \quad $ - {\cal K}^4_{\s(3452)} + {\cal K}^4_{\s(4352)} + {\cal K}^4_{\s(5342)} - {\cal K}^4_{\s(5432)} $\cr
\+ \hfill $ n[ 4_\s  5_\s ,7,1,6, 2_\s  3_\s ]$ & $= + {\cal K}^1_{\s(4523)} - {\cal K}^1_{\s(4532)} - {\cal K}^1_{\s(5423)} + {\cal K}^1_{\s(5432)} $ \cr
\+ & \quad $ - {\cal K}^3_{\s(4523)} + {\cal K}^3_{\s(4532)} + {\cal K}^3_{\s(5423)} - {\cal K}^3_{\s(5432)} $\cr     
\+ \hfill $ n[ 5_\s 7, 4_\s ,1,6, 2_\s  3_\s ]$ & $= + {\cal K}^1_{\s(5423)} - {\cal K}^1_{\s(5432)} + {\cal K}^2_{\s(4523)} - {\cal K}^2_{\s(4532)} $ \cr
\+ & \quad $ + {\cal K}^2_{\s(5423)} - {\cal K}^2_{\s(5432)} + {\cal K}^3_{\s(4523)} - {\cal K}^3_{\s(4532)} $\cr
\+ \hfill $ n[67,1, 2_\s , 3_\s , 4_\s  5_\s ]$ & $= + {\cal K}^5_{\s(2345)} - {\cal K}^5_{\s(2354)} - {\cal K}^5_{\s(2453)} + {\cal K}^5_{\s(2543)} $ \cr
\+ & \quad $ - {\cal K}^5_{\s(3452)} + {\cal K}^5_{\s(3542)} + {\cal K}^5_{\s(4532)} - {\cal K}^5_{\s(5432)} $\cr
\+ \hfill $ n[71,6, 2_\s  3_\s , 4_\s  5_\s ]$  & $= - {\cal K}^1_{\s(2345)} + {\cal K}^1_{\s(2354)} + {\cal K}^1_{\s(3245)} - {\cal K}^1_{\s(3254)} $ \cr
\+ & \quad $ + {\cal K}^1_{\s(4523)} - {\cal K}^1_{\s(4532)} - {\cal K}^1_{\s(5423)} + {\cal K}^1_{\s(5432)} $\cr
\+ \hfill $ n[1 2_\s ,6, 3_\s , 4_\s , 5_\s 7]$ & $= + {\cal K}^2_{\s(2543)} + {\cal K}^3_{\s(2354)} + {\cal K}^3_{\s(2453)} + {\cal K}^3_{\s(2543)} $ \cr
\+ & \quad $ + {\cal K}^4_{\s(2345)} + {\cal K}^4_{\s(2354)} + {\cal K}^4_{\s(2453)} + {\cal K}^5_{\s(2345)} $\cr
\+ \hfill $  n[1 2_\s ,6, 3_\s ,7, 4_\s  5_\s ]$ & $= + {\cal K}^2_{\s(2453)} - {\cal K}^2_{\s(2543)} + {\cal K}^3_{\s(2345)} - {\cal K}^3_{\s(2354)} $ \cr
\+ & \quad $ - {\cal K}^4_{\s(2453)} + {\cal K}^4_{\s(2543)} - {\cal K}^5_{\s(2345)} + {\cal K}^5_{\s(2354)} $\cr
\+ \hfill $  n[ 2_\s  3_\s ,1,6, 4_\s , 5_\s 7]$ & $= - {\cal K}^3_{\s(2354)} + {\cal K}^3_{\s(3254)} - {\cal K}^4_{\s(2345)} - {\cal K}^4_{\s(2354)} $ \cr
\+ & \quad $ + {\cal K}^4_{\s(3245)} + {\cal K}^4_{\s(3254)} - {\cal K}^5_{\s(2345)} + {\cal K}^5_{\s(3245)} $\cr
\+ \hfill $  n[ 2_\s  3_\s , 4_\s ,6,1, 5_\s 7]$ & $= + {\cal K}^1_{\s(5234)} - {\cal K}^1_{\s(5324)} - {\cal K}^1_{\s(5423)} + {\cal K}^1_{\s(5432)} $ \cr
\+ & \quad $ + {\cal K}^2_{\s(5234)} - {\cal K}^2_{\s(5324)} - {\cal K}^2_{\s(5423)} + {\cal K}^2_{\s(5432)} $\cr
\+ \hfill $  n[ 2_\s 6,1, 3_\s  4_\s , 5_\s 7]  = - {\cal K}^1_{\s(5342)} + {\cal K}^1_{\s(5432)} - {\cal K}^2_{\s(5342)} + {\cal K}^2_{\s(5432)} $ \cr
\+ & \quad $ + {\cal K}^3_{\s(3452)} - {\cal K}^3_{\s(4352)} + {\cal K}^4_{\s(3452)} - {\cal K}^4_{\s(4352)} $\cr
\+ \hfill $  n[ 3_\s  4_\s , 5_\s ,7,6,1 2_\s ]$ & $= - {\cal K}^2_{\s(2345)} + {\cal K}^2_{\s(2435)} + {\cal K}^2_{\s(2534)} - {\cal K}^2_{\s(2543)} $ \cr
\+ & \quad $ - {\cal K}^5_{\s(2345)} + {\cal K}^5_{\s(2435)} + {\cal K}^5_{\s(2534)} - {\cal K}^5_{\s(2543)} $\cr
\+ \hfill $  n[ 5_\s 7,6,1, 2_\s , 3_\s  4_\s ]$ & $= - {\cal K}^4_{\s(2345)} + {\cal K}^4_{\s(2435)} + {\cal K}^4_{\s(3425)} - {\cal K}^4_{\s(4325)} $ \cr
\+ & \quad $ - {\cal K}^5_{\s(2345)} + {\cal K}^5_{\s(2435)} + {\cal K}^5_{\s(3425)} - {\cal K}^5_{\s(4325)} $\cr
\+ \hfill $  n[ 2_\s 6,1, 3_\s , 4_\s , 5_\s 7]$ & $= + {\cal K}^1_{\s(5432)} + {\cal K}^2_{\s(3542)} + {\cal K}^2_{\s(4532)} + {\cal K}^2_{\s(5432)} $ \cr
\+ & \quad $ + {\cal K}^3_{\s(3452)} + {\cal K}^3_{\s(3542)} + {\cal K}^3_{\s(4532)} + {\cal K}^4_{\s(3452)} $\cr
%
\+ \hfill $ n[ 4_\s  5_\s ,7,16, 2_\s  3_\s ]$ & $= + {\cal K}^1_{\s(4523)} - {\cal K}^1_{\s(4532)} - {\cal K}^1_{\s(5423)} + {\cal K}^1_{\s(5432)} - {\cal K}^5_{\s(3245)} + {\cal K}^5_{\s(3254)} $ \cr
\+ & \quad $- {\cal K}^3_{\s(2345)} + {\cal K}^3_{\s(2354)} + {\cal K}^3_{\s(3245)} - {\cal K}^3_{\s(3254)} - {\cal K}^3_{\s(4523)} + {\cal K}^3_{\s(4532)}$ \cr
\+ & \quad $ + {\cal K}^3_{\s(5423)} - {\cal K}^3_{\s(5432)} + {\cal K}^5_{\s(2345)} - {\cal K}^5_{\s(2354)} $ \cr
\+ \hfill $ n[16, 2_\s , 3_\s  4_\s , 5_\s 7]$  & $= - {\cal K}^1_{\s(5342)} + {\cal K}^1_{\s(5432)} - {\cal K}^2_{\s(2534)} + {\cal K}^2_{\s(2543)} + {\cal K}^5_{\s(2345)} - {\cal K}^5_{\s(2435)}$ \cr
\+ & \quad $- {\cal K}^2_{\s(5342)} + {\cal K}^2_{\s(5432)} - {\cal K}^3_{\s(2534)} + {\cal K}^3_{\s(2543)} + {\cal K}^3_{\s(3452)} - {\cal K}^3_{\s(4352)} $ \cr
\+ & \quad $+ {\cal K}^4_{\s(2345)} - {\cal K}^4_{\s(2435)} + {\cal K}^4_{\s(3452)} - {\cal K}^4_{\s(4352)}  $\cr
\+ \hfill $  n[16,7, 2_\s  3_\s , 4_\s  5_\s ]$ & $= + {\cal K}^1_{\s(2345)} - {\cal K}^1_{\s(2354)} - {\cal K}^1_{\s(3245)} + {\cal K}^1_{\s(3254)} - {\cal K}^5_{\s(5423)} + {\cal K}^5_{\s(5432)} $ \cr
\+ & \quad $- {\cal K}^1_{\s(4523)} + {\cal K}^1_{\s(4532)} + {\cal K}^1_{\s(5423)} - {\cal K}^1_{\s(5432)} - {\cal K}^5_{\s(2345)} + {\cal K}^5_{\s(2354)} $ \cr
\+ & \quad $+ {\cal K}^5_{\s(3245)} - {\cal K}^5_{\s(3254)} + {\cal K}^5_{\s(4523)} - {\cal K}^5_{\s(4532)}  $ \cr
\+ \hfill $ n[ 2_\s  3_\s , 4_\s , 5_\s ,7,16]$ & $= - {\cal K}^1_{\s(2345)} + {\cal K}^1_{\s(3245)} + {\cal K}^1_{\s(4235)} - {\cal K}^1_{\s(4325)} + {\cal K}^5_{\s(5423)} - {\cal K}^5_{\s(5432)} $ \cr
\+ & \quad $+ {\cal K}^1_{\s(5234)} - {\cal K}^1_{\s(5324)} - {\cal K}^1_{\s(5423)} + {\cal K}^1_{\s(5432)} + {\cal K}^5_{\s(2345)} - {\cal K}^5_{\s(3245)} $ \cr
\+ & \quad $- {\cal K}^5_{\s(4235)} + {\cal K}^5_{\s(4325)} - {\cal K}^5_{\s(5234)} + {\cal K}^5_{\s(5324)} $ \cr
\+ \hfill $ n[ 2_\s  3_\s , 4_\s , 5_\s 7,16]$ & $= + {\cal K}^1_{\s(5234)} - {\cal K}^1_{\s(5324)} - {\cal K}^1_{\s(5423)} + {\cal K}^1_{\s(5432)} - {\cal K}^5_{\s(4235)} + {\cal K}^5_{\s(4325)}$ \cr
\+ & \quad $+ {\cal K}^2_{\s(5234)} - {\cal K}^2_{\s(5324)} - {\cal K}^2_{\s(5423)} + {\cal K}^2_{\s(5432)} + {\cal K}^4_{\s(2345)} - {\cal K}^4_{\s(3245)}$ \cr
\+ & \quad $ - {\cal K}^4_{\s(4235)} + {\cal K}^4_{\s(4325)} + {\cal K}^5_{\s(2345)} - {\cal K}^5_{\s(3245)}  $\cr
\+ \hfill $ n[16, 2_\s , 3_\s , 4_\s , 5_\s 7]$ & $= + {\cal K}^1_{\s(5432)} + {\cal K}^2_{\s(2543)} + {\cal K}^2_{\s(3542)} + {\cal K}^2_{\s(4532)} + {\cal K}^4_{\s(3452)} + {\cal K}^5_{\s(2345)}$ \cr
\+ & \quad $+ {\cal K}^2_{\s(5432)} + {\cal K}^3_{\s(2354)} + {\cal K}^3_{\s(2453)} + {\cal K}^3_{\s(2543)} + {\cal K}^3_{\s(3452)} + {\cal K}^3_{\s(3542)} $ \cr
\+ & \quad $+ {\cal K}^3_{\s(4532)} + {\cal K}^4_{\s(2345)} + {\cal K}^4_{\s(2354)} + {\cal K}^4_{\s(2453)}  $\cr
\+ \hfill $  n[16, 2_\s , 3_\s ,7, 4_\s  5_\s ]$ & $= + {\cal K}^1_{\s(4532)} - {\cal K}^1_{\s(5432)} + {\cal K}^2_{\s(2453)} - {\cal K}^2_{\s(2543)} - {\cal K}^5_{\s(2345)} + {\cal K}^5_{\s(2354)}$ \cr
\+ & \quad $+ {\cal K}^2_{\s(3452)} - {\cal K}^2_{\s(3542)} + {\cal K}^3_{\s(2345)} - {\cal K}^3_{\s(2354)} - {\cal K}^3_{\s(4532)} + {\cal K}^3_{\s(5432)} $ \cr
\+ & \quad $- {\cal K}^4_{\s(2453)} + {\cal K}^4_{\s(2543)} - {\cal K}^4_{\s(3452)} + {\cal K}^4_{\s(3542)}  $\cr
\+ \hfill $  n[ 3_\s  4_\s , 5_\s ,7, 2_\s ,16]$ & $= - {\cal K}^1_{\s(3452)} + {\cal K}^1_{\s(4352)} + {\cal K}^1_{\s(5342)} - {\cal K}^1_{\s(5432)} + {\cal K}^5_{\s(2534)} - {\cal K}^5_{\s(2543)}$\cr
\+ & \quad $- {\cal K}^2_{\s(2345)} + {\cal K}^2_{\s(2435)} + {\cal K}^2_{\s(2534)} - {\cal K}^2_{\s(2543)} - {\cal K}^4_{\s(3452)} + {\cal K}^4_{\s(4352)} $ \cr
\+ & \quad $+ {\cal K}^4_{\s(5342)} - {\cal K}^4_{\s(5432)} - {\cal K}^5_{\s(2345)} + {\cal K}^5_{\s(2435)}  $\cr
\+ \hfill $  n[ 5_\s 7, 4_\s ,16, 2_\s  3_\s ]$ & $= + {\cal K}^1_{\s(5423)} - {\cal K}^1_{\s(5432)} + {\cal K}^2_{\s(4523)} - {\cal K}^2_{\s(4532)} - {\cal K}^5_{\s(2345)} + {\cal K}^5_{\s(3245)}$ \cr
\+ & \quad $+ {\cal K}^2_{\s(5423)} - {\cal K}^2_{\s(5432)} - {\cal K}^3_{\s(2354)} + {\cal K}^3_{\s(3254)} + {\cal K}^3_{\s(4523)} - {\cal K}^3_{\s(4532)} $ \cr
\+ & \quad $ - {\cal K}^4_{\s(2345)} - {\cal K}^4_{\s(2354)} + {\cal K}^4_{\s(3245)} + {\cal K}^4_{\s(3254)}  $ \hskip 2.4cm \sevenpa \cr
They follow from comparing the field theory limit of ${\cal A}_7^{{\rm string}}(1,2_\rho,3_\rho,4_\rho,5_\rho,6_\rho,7)$ with
$$
{\cal A}_7(1,2_\rho,3_\rho,4_\rho,5_\rho,6_\rho,7) = { n \left[ 12_\rho,3_\rho, 4_\rho 5_\rho, 6_\rho 7  \right]
 \over s_{12_\rho} s_{12_\rho 3_\rho } s_{4_\rho 5_\rho}s_{ 6_\rho 7} } -{ n \left[ 2_\rho3_\rho,1, 4_\rho 5_\rho, 6_\rho 7  \right]
 \over s_{2_\rho 3_\rho} s_{12_\rho 3_\rho } s_{4_\rho 5_\rho}s_{ 6_\rho 7} }
 $$ 
$$
+ {n[12_\rho,3_\rho,4_\rho,5_\rho, 6_\rho 7] \over s_{12_\rho} s_{12_\rho 3_\rho} s_{5_\rho 6_\rho 7} s_{6_\rho 7}} - {n[2_\rho3_\rho,1,4_\rho,5_\rho, 6_\rho 7] \over s_{2_\rho 3_\rho} s_{12_\rho 3_\rho} s_{5_\rho 6_\rho 7} s_{6_\rho 7}}- {n[12_\rho,3_\rho,4_\rho,7,5_\rho 6_\rho ] \over s_{12_\rho} s_{12_\rho 3_\rho} s_{5_\rho 6_\rho 7} s_{5_\rho 6_\rho }} 
$$
$$
+ {n[2_\rho 3_\rho,1,4_\rho,7,5_\rho 6_\rho] \over s_{2_\rho 3_\rho} s_{12_\rho 3_\rho} s_{5_\rho 6_\rho 7} s_{5_\rho 6_\rho }}+
{\rm cyclic}(1,2_\rho,3_\rho,4_\rho,5_\rho,6_\rho,7) 
$$

\appendix{C}{Component expressions at four-point}

This appendix gives an example how the supersymmetric expressions for BCJ numerators decompose in components. In general, component expansion spoils the simplicity of the pure spinor superspace results; but they
can be done \PSS. At five point level, for instance, the innocent-looking numerator $\langle T_{12} T_{34} V_5 \rangle$ 
contributes $\sim 100$ terms to the five gluon amplitude. That is why we give no more than the four point 
building blocks $n_s = \langle T_{12} V_3 V_4 \rangle = \langle V_1 V_2 T_{34} \rangle$.

There are four inequivalent bose-fermi combinations to consider, namely 
\eqn \bosefermi{
(1,2,3,4) \in \big\{ (b,b,b,b), (f,f,b,b),(b,f,b,f),(f,f,f,f) \bigr\}.
}
The numerator then evaluates \PSS\ to
$$ 2880 \; n_s \Big|_{bbbb} =  
          -  (k^1 \cdot e^2) (k^1 \cdot e^4) (e^1 \cdot e^3) 
          -  (k^1 \cdot e^2) (k^2 \cdot e^4) (e^1 \cdot e^3) $$
       $$ +  (k^1 \cdot e^2) (k^4 \cdot e^1) (e^3 \cdot e^4) 
        -  (k^1 \cdot e^2) (k^4 \cdot e^3) (e^1 \cdot e^4) 
        +  (k^1 \cdot e^3) (k^1 \cdot e^4) (e^1 \cdot e^2) $$
       $$ +  (k^1 \cdot e^3) (k^2 \cdot e^4) (e^1 \cdot e^2) 
        +  (k^1 \cdot e^4) (k^2 \cdot e^1) (e^2 \cdot e^3) 
        +  (k^1 \cdot e^4) (k^4 \cdot e^3) (e^1 \cdot e^2) $$
       $$ +  (k^2 \cdot e^1) (k^2 \cdot e^4) (e^2 \cdot e^3) 
        -  (k^2 \cdot e^1) (k^4 \cdot e^2) (e^3 \cdot e^4) 
        +  (k^2 \cdot e^1) (k^4 \cdot e^3) (e^2 \cdot e^4) $$
       $$ +  {1 \over 4}(s_{14} - s_{13}) (e^1 \cdot e^2) (e^3 \cdot e^4) 
        +  {s_{12} \over 4}\big[ (e^1 \cdot e^4) (e^2 \cdot e^3)
        -  (e^1 \cdot e^3) (e^2 \cdot e^4) \big] $$       
$$2880\; n_s \Big|_{ffbb}   =   (\chi^1 \g^{e^3} \chi^2) (k^1 \cdot e^4) 
        -  (\chi^1 \g^{k^4} \chi^2) (e^3 \cdot e^4)
        +  (\chi^1 \g^{e^3} \chi^2) (k^2 \cdot e^4) 
        +  (\chi^1 \g^{e^4} \chi^2) (k^4 \cdot e^3) $$
$$2880\;  n_s \Big|_{bfbf}  =  
         {1 \over 2}\big[ (\chi^2 \g^{k^1e^1e^3} \chi^4)
        +  (\chi^2 \g^{k^1} \chi^4) (e^1 \cdot e^3) 
        -   (\chi^2 \g^{e^1} \chi^4) (k^1 \cdot e^3)\big]
        -  (\chi^2 \g^{e^3} \chi^4) (k^2 \cdot e^1) $$
$$2880\;  n_s \Big|_{ffff}  =  (\chi^1 \g^{m} \chi^2) (\chi^3 \g_{m} \chi^4) $$
where the SCHOONSCHIP notation has been used, i.e., $ (\chi^1\g^m \chi^3)e^2_m \equiv (\chi^1\g^{e^2}\chi^3)$ as well as $(\chi^2 \g^{mnp} \chi^4) k^1_m e^1_n e^3_p \equiv (\chi^2 \g^{k^1e^1e^3} \chi^4)$.
The zero mode integrations involved in the four-point calculations have been performed 
in \tsimpis\ for the first time.

\appendix{D}{Spinor helicity evaluation at five-point}

In this appendix, we will express the BCJ numerators of a five gluon MHV amplitude in terms of spinor helicity variables. 
This is meant as a sample calculation showing the relevance of our methods for $D=4$ physics.

Doing so requires running the program \PSS\ for expanding $\langle T_{ij} T_{kl} V_m \rangle$ and 
$\langle T_{ijk} V_l V_m \rangle$ in superfield components, discarding all fermionic contributions. 
As we have emphasized in the previous appendix, evaluation in components spoils the compactness of the 
superspace expressions: Each of the brackets above contain several terms with products of 
polarization vectors and momenta before plugging in the helicity specific spinor products.

If the helicities
of the gluons are $(--+++)$ we use the following
conventions,
\eqn\hel{
e^I_{\a{\dot \a}} = \sqrt{2}{\psi^I_\a {\bar \chi}^I_{\dot \a}\over [{\bar \psi}^I{\bar \chi}^I]}, \quad I=1,2, \quad
e^J_{{\dot \b}\b} = \sqrt{2}{{\bar \psi}^J_{\dot \b}  \chi^J_\b \over \langle  \chi^J \psi^J \rangle}, \quad J=3,4,5
}
where $\langle \psi \chi\rangle = \psi^\a \chi_\a = \e^{\a\b}\psi_\b\chi_\a$ and
$[{\bar \psi}{\bar \chi}] = {\bar \psi}_{\dot \a}{\bar \chi}^{\dot \a} 
= \e_{{\dot \a}{\dot \b}}{\bar \psi}_{\dot \b}{\bar \chi}^{\dot \a}$
are the spinor products and $\langle ij\rangle [ij] = -2 s_{ij}$.
For the specific choice $(2,1,1,1,1)$ of reference momenta $\chi^J_\alpha,\bar \chi^I_{\dot \alpha}$ they imply
$$
 (e^1\cdot e^3) =
 (e^1\cdot e^4) =
 (e^1\cdot e^5) = 
 (e^3\cdot e^4) =
 (e^3\cdot e^5) = 
 (e^4\cdot e^5) = 0
$$
$$
 (k^2\cdot e^1) = 
        (k^1\cdot e^2) =
        (k^1\cdot e^3) = 
        (k^1\cdot e^4) =
        (k^1\cdot e^5) = 0
$$
and we quickly obtain the following basis kinematics (dropping an overall numerical coefficient)
\eqn \MHV{ \eqalign{
{\cal K}^{1}_{(23)} &= \frac{ \langle 12 \rangle^3 [25]^2 [43] }{  [12] \langle 13 \rangle \langle 14 \rangle } , \ \ \
{\cal K}^{1}_{(32)} = \frac{ \langle 12 \rangle^3 [24] [25] [53] }{  [12] \langle 13 \rangle \langle 14 \rangle } , \ \ \ {\cal K}^{2}_{(23)} = 0 \cr
{\cal K}^{2}_{(32)} &= \frac{ \langle 12 \rangle^3 [23] [24] [35] }{  [12] \langle 14 \rangle \langle 51 \rangle } , \ \ \ {\cal K}^3_{(23)} = 0 , \ \ \ {\cal K}^{3}_{(32)} = \frac{ \langle 12 \rangle^3 [23]^2 [45] }{ [12] \langle 14 \rangle \langle 15 \rangle }  
}}
which translates into BCJ numerators $n_1 = n_3 = n_{12} = 0$ and
\eqn \MHVnum{ \eqalign{
(n_2,&n_4,n_5,n_6,n_7,n_8) = \frac{ \langle 12 \rangle^3 }{ [12] } \cr
&\times \left( \frac{[23 ][25] [45] }{ \langle 13 \rangle \langle 14 \rangle }, \frac{[23 ]^2 [45] }{ \langle 14 \rangle \langle 51 \rangle }, \frac{[25]^2 [43] }{ \langle 13 \rangle \langle 14 \rangle }, \frac{[24 ][25] [34] }{ \langle 13 \rangle \langle 15 \rangle } , \frac{[23 ][24] [45] }{ \langle 13 \rangle \langle 51 \rangle }, \frac{[25]^2 [43] }{ \langle 13 \rangle \langle 14 \rangle }\right) \cr
(n_9,n_{10},&n_{11},n_{13},n_{14},n_{15}) = \frac{ \langle 12 \rangle^3 }{ [12] } \cr
&\times \left( \frac{[23 ][25] [34] }{ \langle 14 \rangle \langle 51 \rangle }, \frac{[23 ] [24] [35] }{ \langle 14 \rangle \langle 51 \rangle }, \frac{[24][25] [35] }{ \langle 13 \rangle \langle 14 \rangle }, \frac{[24 ]^2[35] }{ \langle 13 \rangle \langle 51 \rangle } , \frac{[24]^2 [35] }{ \langle 13 \rangle \langle 15 \rangle }, \frac{[23]^2 [45] }{ \langle 14 \rangle \langle 15 \rangle }\right)
}}
They can be easily checked to reproduce the Parke-Taylor formula \refs{\ParkeTaylor,\BG}.

All the nonlocalities $\sim [12]^{-1},\langle 13 \rangle^{-1},\langle 14 \rangle^{-1}, \langle 15 \rangle^{-1}$ are
spurious and arise from the reference momentum dependent denominators of \hel. As we have emphasized, all the $n_i$ in this work are local in 
any spacetime dimension.

\listrefs

\end